\documentclass[11pt]{article}

\usepackage[dvips]{graphicx}

\newcommand{\re}{\mathrm{Re}\,}
\newcommand{\im}{\mathrm{Im}\,}

\begin{document}

\newcommand{\tvec}[1]{\mbox{\boldmath{$#1$}}}
\newcommand{\svec}[1]{\mbox{\boldmath{$\scriptstyle #1$}}}

\newcommand{\half}{{\textstyle\frac{1}{2}}}

\begin{flushright}
DESY 05-037 \\
hep-ph/0503023 \\
\end{flushright}

\begin{center}
\vskip 4.0\baselineskip
{\LARGE \bf On the analysis of lepton scattering \\[0.4em]
  on longitudinally or transversely polarized protons}

\vskip 8.0\baselineskip
M.~Diehl \\[0.5\baselineskip]
\textit{Deutsches Elektronen-Synchroton DESY, 22603 Hamburg, Germany}
\\[1.5\baselineskip] 
S.~Sapeta \\[0.5\baselineskip]
\textit{M.~Smoluchowski Institute of Physics, Jagellonian University,
  Cracow, Poland}
\vskip 4.0\baselineskip
\textbf{Abstract}\\[0.5\baselineskip]
\parbox{0.9\textwidth}{We discuss polarized lepton-proton scattering
  with special emphasis on the difference between target polarization
  defined relative to the lepton beam or to the virtual photon
  direction.  In particular, this difference influences azimuthal
  distributions in the final state.  We provide a general framework of
  analysis and apply it to the specific cases of semi-inclusive deep
  inelastic scattering, of exclusive meson production, and of deeply
  virtual Compton scattering.}
\end{center}

\vfill

\newpage

\tableofcontents

\newpage


\section{Introduction}
\label{sec:intro}

Measurements of deep inelastic scattering on a polarized nucleon are
an essential source of information in spin physics.  The inclusive
spin dependent structure functions $g_1$ and $g_2$ have become
textbook material, and present-day experiments investigate selected
final states that give access to a wealth of information about the
role of spin in the internal structure of the nucleon.  In
semi-inclusive deep inelastic scattering (SIDIS) for instance, the
Collins effect \cite{Collins:1992kk} provides an opportunity to access
the transversity distribution of quarks, and the Sivers effect
\cite{Sivers:1989cc} reveals the subtle role of gluon rescattering in
QCD dynamics \cite{Brodsky:2002cx}.  In exclusive channels like meson
electroproduction and deeply virtual Compton scattering (DVCS), target
polarization allows one to separate generalized parton distributions
with different spin dependence.  In particular, the transverse target
spin asymmetry for appropriate final states \cite{Goeke:2001tz} is
sensitive to the helicity-flip distribution $E$, which carries
information about the orbital angular momentum of quarks in the
nucleon \cite{Ji:1996ek}.

In experiment, the target polarization usually is longitudinal or
transverse with respect to the \emph{lepton beam} direction.  For the
strong-interaction part of the reaction, i.e., the $\gamma^* p$
subprocess, longitudinal and transverse polarization with respect to
the \emph{virtual photon} momentum is however a more natural basis.
The conversion between the two sets of polarization states is simple
and well known for a target polarized longitudinally with respect to
the lepton beam, whereas for transverse polarization the
transformation is more involved.  In the present contribution, we give
a general framework to analyze transverse and longitudinal
polarization data, both for semi-inclusive and for exclusive
processes.

The outline of this paper is as follows.  In Sect.~\ref{sec:trafo} we
give the general transformation between target polarization
longitudinal or transverse with respect to either the lepton beam or
the virtual photon direction.  In Sects.~\ref{sec:ep2gp} and
\ref{sec:dihadron} we derive and discuss the general expression of the
polarized lepton-proton cross section in terms of cross sections and
interference terms at the $\gamma^* p$ level.  We apply these results
to the specific cases of SIDIS and exclusive meson production in
Sects.~\ref{sec:sidis} and \ref{sec:mesons}.  In
Sect.~\ref{sec:positive} we derive positivity bounds and show how they
may help one to separate contributions from longitudinal and
transverse photons in the cross section.  The special case of DVCS is
discussed in Sect.~\ref{sec:dvcs}, and we summarize our results in
Sect.~\ref{sec:sum}.  Some additional material is given in three
appendices.


\section{Transformation of the target spin}
\label{sec:trafo}

\begin{figure}[p]
\begin{center}
\leavevmode
\includegraphics[width=0.8\textwidth]{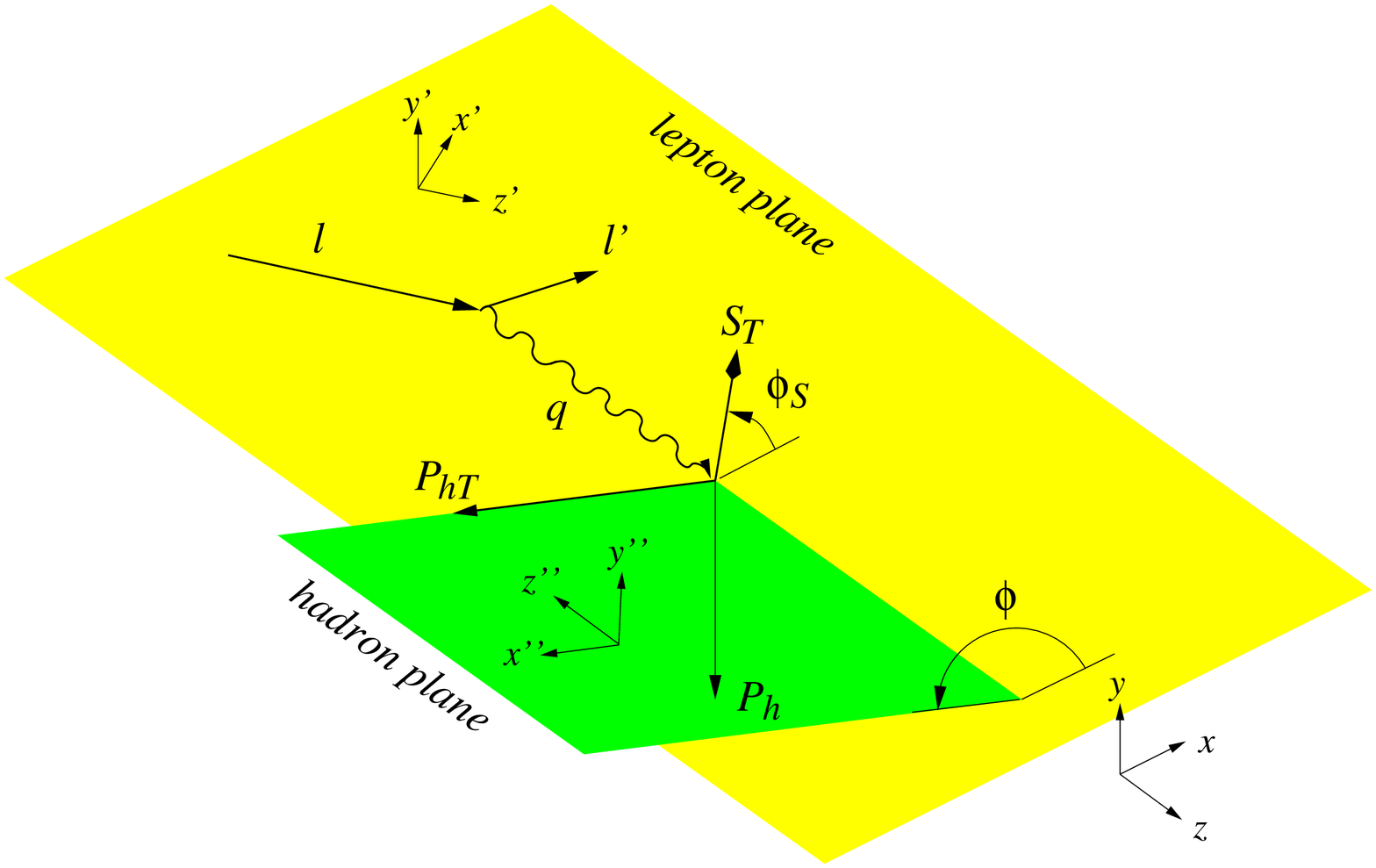}
\end{center}
\caption{\label{fig:coordinates} Kinematics of the process
  (\protect\ref{gen-proc}) in the target rest frame.  $\tvec{P}_{hT}$
  and $\tvec{S}_T$ respectively are the components of $\tvec{P}_h$ and
  $\tvec{S}$ perpendicular to $\tvec{q}$.  (The target spin vector
  $\tvec{S}$ is not shown.)  $\phi$ and $\phi_S$ respectively are the
  azimuthal angles of $\tvec{P}_h$ and $\tvec{S}$ in the coordinate
  system with axes $x$, $y$, $z$, in accordance with the Trento
  conventions \cite{Bacchetta:2004jz}.}
\end{figure}

We consider lepton-proton scattering processes of the form
\begin{equation}
  \label{gen-proc}
\ell(l) + p(P) \to \ell(l') + h(P_h) + X(P')
\end{equation}
with four-momenta given in parentheses.  $\ell$ denotes the lepton,
$p$ the target proton, and $h$ a produced hadron.  $X$ can be an
inclusive system of hadrons as in SIDIS, or a single hadron as in
exclusive processes.  The virtual photon radiated by the lepton has
momentum $q= l-l'$.  We use the conventional kinematical variables for
deep inelastic processes, $Q^2= -q^2$, $x_B= Q^2 /(2 P\cdot q)$,
$y=(P\cdot q) /(P\cdot l)$, and the azimuthal angle $\phi$ between the
hadron and lepton planes as shown in Fig.~\ref{fig:coordinates}.  Our
discussion in this section also covers the case of virtual Compton
scattering, where $h$ is a real photon, as well as processes where $h$
is a system of several particles.  In this section we do not make any
kinematical approximations, except for neglecting the lepton mass.

\begin{figure}[p]
\begin{center}
\leavevmode
\includegraphics[width=0.33\textwidth]{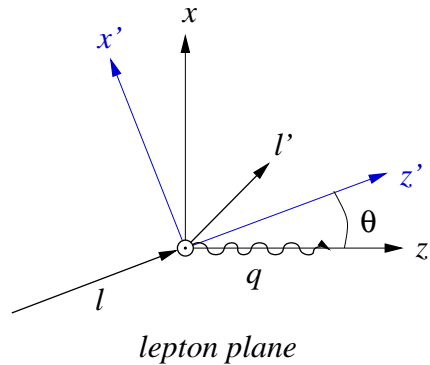}
\end{center}
\caption{\label{fig:lepton-plane} The lepton plane in the target rest
  frame.  The $y$ and $y'$ axes coincide and point out of the paper
  plane.}
\end{figure}

To transform between the different target polarization states, we find
it useful to introduce two coordinate systems $C$ and $C'$ in the
target rest frame, with respective axes $x$, $y$, $z$ and $x'$, $y'$,
$z'$ as shown in Figs.~\ref{fig:coordinates} and
\ref{fig:lepton-plane}.  The $z$ axis points along $\tvec{q}$, whereas
the $z'$ axis points along $\tvec{l}$.  The $x$ axis and the $x'$ axis
are chosen such that $\tvec{l}'$ lies in the $x$-$z$ and the $x'$-$z'$
plane and has a positive $x$ and $x'$ component.  The $y$ and $y'$
axes coincide.  The two coordinate systems $C$ and $C'$ are related
via a rotation about the $y$ axis by the angle $\theta$ between
$\tvec{q}$ and $\tvec{l}$.  In terms of invariants we have
\begin{equation}
  \label{theta}
\sin\theta = \gamma\, 
  \sqrt{\frac{1-y- \frac{1}{4} y^2 \gamma^2}{1+\gamma^2}} , 
\qquad\qquad
\gamma = 2 x_B M_p/Q ,
\end{equation}
where $M_p$ is the proton mass.  In deep inelastic kinematics $\gamma$
is small, and so is $\sin\theta \approx \gamma \sqrt{1-y
\rule{0pt}{0.75em}}$.  Note for instance that $\gamma^2$ is the
parameter controlling the size of target mass corrections in inclusive
DIS~\cite{Nachtmann:1973mr}.

We parameterize the target spin vector $\tvec{S}$ in the two
coordinate systems by
\begin{equation}
\tvec{S} \,\stackrel{C}{=}\,
\left( \begin{array}{c}
S_T \cos\phi_S \\ S_T \sin\phi_S \\ -S_L
\end{array} \right) ,
\qquad \qquad
\tvec{S} \,\stackrel{C'}{=}\,
\left( \begin{array}{c}
P_T \cos\psi \\ P_T \sin\psi \\ -P_L
\end{array} \right) ,
\end{equation}
so that $P_L$, $P_T$ specify longitudinal and transverse polarization
relative to the lepton beam direction, and $S_L$, $S_T$ longitudinal
and transverse polarization relative to the virtual photon direction.
Likewise, $\psi$ is the azimuthal angle of the target spin around the
lepton beam direction, whereas $\phi_S$ is the corresponding azimuthal
angle around the virtual photon direction.  $P_L$ and $S_L$ are
between $-1$ and~$1$, and $P_T$ and $S_T$ are between 0 and 1.  The
sign convention for the longitudinal case is such that $P_L= +1$ and
$S_L= +1$ correspond to a right-handed proton in the $\ell p$ and
$\gamma^* p$ center of mass, respectively.  The values of $P_L$ and
$P_T$ are determined by the experimental setup, whereas $S_L$ and
$S_T$ depend on the kinematics of an individual event.  The rotation
from $C$ to $C'$ readily gives
\begin{eqnarray}
S_T \cos\phi_S &=& \cos\theta\, P_T \cos\psi - \sin\theta\, P_L ,
\nonumber \\
S_T \sin\phi_S &=& P_T \sin\psi ,
\nonumber \\
S_L            &=& \sin\theta\, P_T \cos\psi + \cos\theta\, P_L .
\end{eqnarray}
We remark that, although we work in the target rest frame, our results
can readily be applied to a polarized $\ell p$ collider, whose
laboratory frame is obtained from the target rest frame by a boost
along the lepton beam momentum.  $P_L$ and $P_T$ then give the
longitudinal and transverse polarization of the proton beam with
respect to the beam axis.

\subsection{Longitudinal polarization with respect to the lepton beam}
\label{sec:long}

We have $P_T=0$, so that 
\begin{equation}
  \label{long-rel}
S_L = \cos\theta\, P_L , \qquad\qquad
S_T \cos\phi_S = - \sin\theta\, P_L , \qquad\qquad
S_T \sin\phi_S = 0 .
\end{equation}
If we allow $S_T$ to be negative, so that $(S_T, \phi_S)$ and $(-S_T,
\phi_S+\pi)$ are equivalent, the second and third relation can be
written more simply as $S_T = - \sin\theta\, P_L$ and $\phi_S=0$.

\subsection{Transverse polarization with respect to the lepton beam}
\label{sec:transv}

With $P_L=0$ we find
\begin{equation}
  \label{transv-rel-prelim}
 S_T \cos\phi_S = \cos\theta\, P_T \cos\psi , \qquad
 S_T \sin\phi_S = P_T \sin\psi , \qquad
 S_L            = \sin\theta\, P_T \cos\psi .
\end{equation}
It turns out that the expression for the cross section in the next
sections are considerably simpler when written in terms of the angle
$\phi_S$ instead of $\psi$.  We can use the relations
(\ref{transv-rel-prelim}) to obtain
\begin{equation}
  \label{phi2psi}
\sin\psi = \frac{\cos\theta\,
           \sin\phi_S}{\sqrt{1-\sin^2\!\theta \, \sin^2\!\phi_S}} ,
\qquad\qquad
\cos\psi = 
    \frac{\cos\phi_S}{\sqrt{1-\sin^2\!\theta \, \sin^2\!\phi_S}}
\end{equation}
and, inserting this into the same relations, finally have
\begin{equation}
  \label{transv-rel}
 S_T = 
  \frac{\cos\theta}{\sqrt{1-\sin^2\!\theta \, \sin^2\!\phi_S}}\, P_T ,
\qquad\qquad
 S_L = \frac{\sin\theta\,
  \cos\phi_S}{\sqrt{1-\sin^2\!\theta \, \sin^2\!\phi_S}}\, P_T .
\end{equation}
The phase space element is however simpler in terms of $\psi$, which
describes the azimuthal distribution of the scattered lepton around
the beam axis, with the reference direction provided by the target
spin.\footnote{In the case where $P_T=0$ one can define $\psi$ as the
azimuthal angle of $\tvec{l}'$ with respect to an arbitrary direction
fixed in space.  The cross section is then of course independent of
this angle.}
Namely, we have
\begin{equation}
  \label{phase-space}
  \frac{d^3 l'}{2 l'^0} = \frac{y}{4x_B}\, dx_B\, dQ^2\, d\psi 
  = \frac{y}{4x_B}\, dx_B\, dQ^2\, d\phi_S\, 
      \frac{\cos\theta}{1 - \sin^2\!\theta\, \sin^2\!\phi_S} .
\end{equation}
The transformation from $d\psi$ to $d\phi_S$ introduces an explicit
$\phi_S$ dependence.  In deep inelastic kinematics, one has however
$d\psi \approx d\phi_S$ up to corrections of order~$\gamma^2$.

\subsection{Cross section and asymmetries}

The dependence of the $\ell p$ cross section on the target
polarization is at most linear in the spin vector~$\tvec{S}$. This
follows from the superposition principle and becomes for instance
explicit in the spin density matrix formalism used in the next
section.  For an unpolarized lepton beam we can therefore write
$d\sigma /(dx_B\, dQ^2\, d\phi\, d\psi) = a_0 + \tvec{S} \cdot
\tvec{a}$, where $a_0$ and $\tvec{a}$ only depend on the four-momenta
of the reaction (\ref{gen-proc}) but not on the target spin.
Expressing the vectors in our coordinate system $C$ we have
\begin{equation}
  \label{Xsect-gen}
\frac{d\sigma}{dx_B\, dQ^2\, d\phi\, d\psi} =
  a_0 + S_T \cos\phi_S\, a_1 + S_T \sin\phi_S\, a_2 - S_L a_3 ,
\end{equation}
where the $a_i$ depend on $x_B$, $y$, $Q^2$ and $\phi$ but not on
$\phi_S$ or $\psi$.  With (\ref{long-rel}) and (\ref{transv-rel}) we
have
\begin{eqnarray}
  \label{Xsect-pol}
\frac{1}{2\pi}\, \frac{d\sigma}{dx_B\, dQ^2\, d\phi} \,\Bigg|_{P_T=0}
  &=& a_0 - P_L \sin\theta\, a_1 - P_L \cos\theta\, a_3 ,
\nonumber \\
\frac{d\sigma}{dx_B\, dQ^2\, d\phi\, d\phi_S} \,\Bigg|_{P_L=0}
  &=& \frac{\cos\theta}{1 - \sin^2\!\theta\, \sin^2\!\phi_S}
\nonumber \\
 &\times& \Bigg[ \, a_0 + P_T\, 
    \frac{\cos\theta \cos\phi_S\, a_1
        + \cos\theta \sin\phi_S\, a_2 
        - \sin\theta \cos\phi_S\, a_3}{\sqrt{1 
              - \sin^2\!\theta\, \sin^2\!\phi_S}} \, \Bigg] ,
\end{eqnarray}
where in the first relation we have integrated over $\psi$ and in the
second one we have used (\ref{phase-space}) to trade $d\psi$ for
$d\phi_S$.

It is often useful to express the spin dependence of a process through
asymmetries.  We define asymmetries for longitudinal and transverse
target polarization with respect to the lepton beam
\begin{eqnarray}
  \label{Aell-def}
A_{UL}^{\ell} &=& 
  \frac{d\sigma(P_L=+1) - d\sigma(P_L=-1)}{
        d\sigma(P_L=+1) + d\sigma(P_L=-1)} \; \Bigg|_{P_T=0} \, ,
\nonumber \\
A_{UT}^{\ell}(\phi_S) &=&
  \frac{d\sigma(\phi_S) - d\sigma(\phi_S+\pi)}{
        d\sigma(\phi_S) + d\sigma(\phi_S+\pi)} \; \Bigg|_{P_T=1, P_L=0}
\end{eqnarray}
in accordance with the Trento conventions \cite{Bacchetta:2004jz}.
The subscript $U$ indicates an unpolarized lepton beam, and for better
legibility we have not displayed the dependence of the cross sections
and asymmetries on other kinematical variables $\phi$, $x_B$, $Q^2$,
etc.  These asymmetries can be directly measured in experiment,
whereas their counterparts for longitudinal and transverse target
polarization with respect to the virtual photon direction
\begin{eqnarray}
  \label{Agamma-def}
A_{UL}^{\gamma^*} &=& 
  \frac{d\sigma(S_L=+1) - d\sigma(S_L=-1)}{
        d\sigma(S_L=+1) + d\sigma(S_L=-1)} \; \Bigg|_{S_T=0} \, ,
\nonumber \\
A_{UT}^{\gamma^*}(\phi_S) &=&
  \frac{d\sigma(\phi_S) - d\sigma(\phi_S+\pi)}{
        d\sigma(\phi_S) + d\sigma(\phi_S+\pi)} \; \Bigg|_{S_T=1, S_L=0}
\end{eqnarray}
are more natural to describe the physics of the $\gamma^* p$
subprocess.  {}From (\ref{Xsect-gen}) and (\ref{Xsect-pol}) we readily
obtain the transformation between the two types of asymmetries,
\begin{eqnarray}
  \label{asy-transf}
A_{UL}^{\ell} &=& 
  \cos\theta\, A_{UL}^{\gamma^*}
  - \sin\theta\, A_{UT}^{\gamma^*}(0) \, ,
\nonumber \\[0.4em]
A_{UT}^{\ell}(\phi_S) &=& 
  \frac{\cos\theta\, A_{UT}^{\gamma^*}(\phi_S) 
      + \sin\theta\, \cos\phi_S\, A_{UL}^{\gamma^*}
  \rule[-0.4em]{0pt}{1em}}{\sqrt{1-\sin^2\!\theta \, \sin^2\!\phi_S}}
\end{eqnarray}
and its inverse
\begin{eqnarray}
  \label{asy-inverse}
A_{UL}^{\gamma^*} &=&
  \cos\theta\, A_{UL}^{\ell} + \sin\theta\, A_{UT}^{\ell}(0) \, ,
\nonumber \\[0.4em]
A_{UT}^{\gamma^*}(\phi_S) &=& 
  \frac{\sqrt{1-\sin^2\!\theta \, \sin^2\!\phi_S}\;
    A_{UT}^{\ell}(\phi_S)}{\cos\theta} 
  - \sin\theta\, \cos\phi_S\, \Big( A_{UL}^{\ell} + \tan\theta\,
                                    A_{UT}^{\ell}(0) \Big)
\nonumber \\[0.8em]
 &=& \cos\phi_S\, \Big( \cos\theta\, A_{UT}^{\ell}(0)
                      - \sin\theta\, A_{UL}^{\ell} \,\Big)
   + \sin\phi_S\, A_{UT}^{\ell}(\half\pi) \, .
\end{eqnarray}
An experiment having both longitudinal and transverse target
polarization can hence uniquely reconstruct the asymmetries
$A_{UL}^{\gamma^*}$ and $A_{UT}^{\gamma^*}(\phi_S)$.  To determine
$A_{UT}^{\ell}(\phi_S)$ at $\phi_S=0$ or $\phi_S=\half\pi$ one can of
course use data for all $\phi_S$, given that
\begin{equation}
  \label{AUT-general}
A_{UT}^{\ell}(\phi_S) = 
\frac{\cos\phi_S\, A_{UT}^{\ell}(0) 
    + \cos\theta\, \sin\phi_S\, A_{UT}^{\ell}(\half\pi)
  \rule[-0.4em]{0pt}{1em}}{\sqrt{1-\sin^2\!\theta \, \sin^2\!\phi_S}}
\end{equation}
according to (\ref{Xsect-pol}).  Notice that the transformations
(\ref{asy-transf}) and (\ref{asy-inverse}) require $\theta$ to be
fixed, which implies that the cross sections in (\ref{Aell-def}) and
(\ref{Agamma-def}) have to be differential in both $x_B$ and $Q^2$
(which also fixes $y$ for a given c.m.\ energy of the $\ell p$
collision).  If the measured cross sections are integrated over wider
bins in $x_B$ and $Q^2$, the transformations can only be done
approximately, with an average value of $\theta$.

Our results generalize straightforwardly to the case of a
longitudinally polarized lepton beam.  The relations (\ref{Xsect-gen})
and (\ref{Xsect-pol}) then hold separately for right- and left-handed
beam polarization with coefficients $a_i^\rightarrow$ and
$a_i^\leftarrow$ .  (Since we neglect the lepton mass, the lepton
helicity is a good quantum number and frame independent.)  Writing
$d\sigma^\rightarrow$ and $d\sigma^\leftarrow$ for the respective
cross section with a right-handed and left-handed lepton beam, we
introduce double spin asymmetries
\begin{eqnarray}
  \label{Adouble-def}
A_{LL}^{\ell} &=& 
  \frac{d\sigma^\rightarrow(P_L=+1) - d\sigma^\rightarrow(P_L=-1)
      - d\sigma^\leftarrow(P_L=+1)  + d\sigma^\leftarrow(P_L=-1)}{
        d\sigma^\rightarrow(P_L=+1) + d\sigma^\rightarrow(P_L=-1)
      + d\sigma^\leftarrow(P_L=+1)  + d\sigma^\leftarrow(P_L=-1)}
  \; \Bigg|_{P_T=0} \, ,
\nonumber \\
A_{LT}^{\ell}(\phi_S) &=&
  \frac{d\sigma^\rightarrow(\phi_S) - d\sigma^\rightarrow(\phi_S+\pi)
      - d\sigma^\leftarrow(\phi_S)  + d\sigma^\leftarrow(\phi_S+\pi)}{
        d\sigma^\rightarrow(\phi_S) + d\sigma^\rightarrow(\phi_S+\pi)
      + d\sigma^\leftarrow(\phi_S)  + d\sigma^\leftarrow(\phi_S+\pi)}
  \; \Bigg|_{P_T=1, P_L=0}
\end{eqnarray}
and their analogs $A_{LL}^{\gamma^*}$ and $A_{LT}^{\gamma^*}(\phi_S)$,
with $(P_L, P_T)$ replaced by $(S_L, S_T)$.  One then has relations
like (\ref{asy-transf}), (\ref{asy-inverse}) and (\ref{AUT-general})
with the subscript $U$ replaced by $L$.


\section{From $\ell p$ to $\gamma^* p$ cross sections}
\label{sec:ep2gp}

In the previous section we have given the transformation between
target polarization defined with respect to either the direction of
$\tvec{l}$ or the direction of $\tvec{q} = \tvec{l} - \tvec{l}'$.  We
have not actually used that $\tvec{q}$ is the momentum of a virtual
photon which is radiated off the lepton beam and absorbed by the
target proton.  We now use this, which will in particular allow us to
make explicit the interplay between the azimuthal angles $\phi$ and
$\phi_S$.  The discussion in this chapter holds for processes like
SIDIS and exclusive meson production, but not for DVCS (see
Sect.~\ref{sec:dvcs}).

Our evaluation of the $\ell p$ cross section closely follows the steps
detailed in Sect.~3 of \cite{Arens:1996xw} for an unpolarized target.
A reader not interested in the derivation may directly go to the
result (\ref{Xsection-master}).  To describe the $\gamma^* p$
subprocess we use a coordinate system $C''$ with axes $x''$, $y''$,
$z''$ as shown in Fig.~\ref{fig:coordinates}.  The $z''$ axis points
opposite to $\tvec{q}$ and the $x''$ axis is chosen such that
$\tvec{P}_h$ lies in the $x''$-$z''$ plane and has a positive $x''$
component.\footnote{We take the $z''$ axis opposite to the $z$ axis of
coordinate system $C$, so that in the $\gamma^* p$ center of mass the
proton moves into the positive $z''$ direction, a choice favored in
many theoretical calculations.}
In this coordinate system the proton spin vector reads
\begin{equation}
\tvec{S} \,\stackrel{C''}{=}\,
\left( \begin{array}{c}
S_T \cos(\phi-\phi_S) \\ S_T \sin(\phi-\phi_S) \\ S_L
\end{array} \right)
\end{equation}
and the spin density matrix of the target \cite{Martin:1970aa} can be
written as
\begin{equation}
  \label{spin-density}
\rho_{ji} \,=\, \frac{1}{2} \Big[ \delta_{ji} 
  + \tvec{S} \cdot \tvec{\sigma}_{ji} \Big]
\;\stackrel{C''}{=}\;  \frac{1}{2} 
\left( \begin{array}{cc}
  1 + S_L & S_T \exp[-i(\phi-\phi_S)] \\
  S_T \exp[i(\phi-\phi_S)] & 1 - S_L
\end{array} \right)
\end{equation}
in a basis of polarization states specified by two-component spinors
\begin{equation}
  \label{two-spinors}
\chi_{+\frac{1}{2}} 
  = \left( \begin{array}{c} 1 \\ 0 \end{array} \right) ,
\qquad\qquad
\chi_{-\frac{1}{2}} 
  = \left( \begin{array}{c} 0 \\ 1 \end{array} \right) .
\end{equation}
These states respectively correspond to definite spin projection
$+\frac{1}{2}$ and $-\frac{1}{2}$ along the $z''$ axis, and to right-
and left-handed proton helicity in the $\gamma^* p$ center of mass.
The components of $\tvec{\sigma}$ in (\ref{spin-density}) are the
Pauli matrices.  As is well known, the cross section can be written as
\begin{equation}
d\sigma(\ell p\to \ell h X) \propto L^{\nu\mu} W_{\mu\nu} \;
  \frac{d^3 l'}{2l'^0}\, \frac{d^3 P_h}{2P_h^0}   ,
\end{equation}
with a proportionality factor depending on $x_B$, $y$ and $Q^2$.  The
leptonic tensor reads
\begin{equation}
L^{\nu\mu} = l'^\nu l^\mu + l^\nu l'^\mu - (l' \cdot l)\, g^{\nu\mu} +
   i P_\ell\, \epsilon^{\nu\mu\alpha\beta} q_\alpha l_\beta
\end{equation}
with the convention $\epsilon_{0123} = 1$ and the lepton beam
polarization $P_\ell$ defined such that $P_\ell=+1$ corresponds to a
purely right-handed and $P_\ell=-1$ to a purely left-handed beam.  The
hadronic tensor is given by
\begin{equation}
  \label{had-tensor}
W_{\mu\nu} = \sum_{ij} \rho_{ji}  \sum_X
    \delta^{(4)}(P'+P_h -P-q) \sum_{\rm spins}
    \langle p(i) | J_\mu(0) | h X \rangle \, 
    \langle h X | J_\nu(0) | p(j) \rangle ,
\end{equation}
where $J_\mu$ is the electromagnetic current.  $\sum_X$ denotes the
integral over the momenta of all hadrons in $X$, and also the sum over
their number if $X$ is an inclusive system.  There are further sums
$\sum_{ij}$ over target spin states $i,j= \pm \frac{1}{2}$ and
$\sum_{\rm spins}$ over all polarizations in the hadronic final state
$h X$.  We now introduce polarization vectors $\epsilon_{m}$ for
definite helicity $m$ of the virtual photon,
\begin{eqnarray}
  \label{photon-pol}
\epsilon_0^\mu &=& \frac{1}{Q \sqrt{1+\gamma^2}}\,
  \Big( q^\mu + \frac{Q^2}{P\cdot q}\, P^\mu \Big) ,
\nonumber \\
\epsilon_{+1} &=&   \frac{1}{\sqrt{2}}\, (0,-1,i,0) , \qquad\qquad
\epsilon_{-1} \;=\: \frac{1}{\sqrt{2}}\, (0,1,i,0) ,
\end{eqnarray}
with $\gamma$ defined in (\ref{theta}) and the components of
$\epsilon_{\pm 1}$ given in coordinate system $C''$.  As shown in
\cite{Arens:1996xw}, the leptonic tensor $L^{\nu\mu}$ can be expressed
as a linear combination of terms $\epsilon^\nu_n \epsilon^{\mu *}_m$.
Up to a global factor the expansion coefficients form the spin density
matrix of the virtual photon.  They depend on $P_\ell$, on $Q^2$, on
the usual ratio of longitudinal and transverse photon flux
\begin{equation}
  \label{eps-def}
\varepsilon
  = \frac{1 - y - \frac{1}{4} y^2 \gamma^2}{
          1 - y + \frac{1}{2} y^2 + \frac{1}{4} y^2 \gamma^2} \, ,
\end{equation}
and on the azimuthal angle $\phi$.\footnote{The polarization vectors in
(\protect\ref{photon-pol}) are identical to those in Eq.~(3.16) of
\protect\cite{Arens:1996xw}, where they are however given in a
different coordinate system.  We also note that the angle $\varphi$ in
\protect\cite{Arens:1996xw} is equal to $-\phi$ used here.}
The contraction $L^{\nu\mu} W_{\mu\nu}$ can then be written in terms
of quantities
\begin{equation}
  \label{sigma-mn}
\rule[-1.2em]{0pt}{3.2em}
\sigma_{mn}
 = \sum_{ij} \rho_{ji}^{\phantom{m}}\, \sigma^{ij}_{mn} 
 \propto \int dt\, dM_X^2\,
    (\epsilon^{\mu *}_m W_{\mu\nu}^{\phantom{*}}\, \epsilon^\nu_n) ,
\end{equation}
where the $x_B$ and $Q^2$ dependent proportionality factor is chosen
such that $\sigma_{mm}$ is the $\gamma^* p$ cross section for photon
helicity $m$ with Hand's convention for the virtual photon flux.  In
(\ref{sigma-mn}) we have integrated over the invariant momentum
transfer $t = (P-P')^2 = (P_h-q)^2$ and over the invariant mass $M_X^2
= P'^2$ of the system $X$.\footnote{The integration over $M_X^2$ is
trivial if $X$ is a single hadron, because then $\sum_X = (2\pi)^{-3}
\int d^3P' /(2 P'^0)$.  Together with $\delta^{(4)}(P'+P_h -P-q)$ this
leaves one delta function constraint in the hadronic tensor
(\ref{had-tensor}).}
The $\sigma^{ij}_{mn}$ are polarized photoabsorption cross sections or
interference terms, given by
\begin{equation}
  \label{sigma-mn-ij}
\sigma^{ij}_{mn}(x_B, Q^2) \propto \int dt\, dM_X^2\,
   \sum_{X} \delta^{(4)}(P'+P_h -P-q) \sum_{\rm spins}
   \Big( \mathcal{A}^i_m \Big)^* \mathcal{A}^j_n
\end{equation}
in terms of the amplitudes $\mathcal{A}^i_m$ for the subprocess
$\gamma^* p\to h X$ with proton polarization $i$ and photon
polarization $m$.  Changing the basis of spin states one can rewrite
interference terms as linear combinations of cross sections, as shown
in App.~\ref{app:inter-cross}.  We have defined our polarization
states for protons and photons in the coordinate system $C''$, whose
axes are specified with reference only to the momenta of the $\gamma^*
p$ process, but \emph{not} to the lepton momenta or to the proton
polarization.  Therefore $\sigma^{ij}_{mn}$ depends on the kinematical
variables $x_B$ and $Q^2$, whereas the dependence on $\varepsilon$ and
$\phi$ is contained in $L^{\nu\mu}$ and the dependence on $S_T$, $S_L$
and $\phi_S$ in $\rho_{ji}$.  {}From hermiticity and parity invariance
we have relations $\sigma_{nm}^{\phantom{*}} = \sigma_{mn}^*$ and
\begin{equation}
  \label{sigma-mn-ij-props}
\rule[-1.2em]{0pt}{3.2em}
\sigma^{ji}_{nm} = (\sigma^{ij}_{mn})^* ,
\qquad\qquad
\sigma^{-i-j}_{-m-n} = (-1)^{m-n-i+j}\, \sigma^{ij}_{mn} 
\end{equation}
with $m,n = 0,+1,-1$ and $i,j = +\frac{1}{2}, -\frac{1}{2}$.  They
imply that $\sigma_{00}^{+-}$, $\sigma_{+-}^{+-}$ and
$\sigma_{+-}^{-+}$ are purely imaginary, whereas other interference
terms have both real and imaginary parts.  Using these relations and
closely following the steps of the derivation in \cite{Arens:1996xw}
we obtain our master formula
\begin{eqnarray}
  \label{Xsection-master}
\lefteqn{
\Bigg[ \frac{\alpha_{\rm em}}{8\pi^3}\, \frac{y^2}{1-\varepsilon}\,
       \frac{1-x_B}{x_B}\, \frac{1}{Q^2} \Bigg]^{-1}
\frac{d\sigma}{dx_B\, dQ^2\, d\phi\, d\psi}
}
\nonumber \\[0.2em]
 &=& \frac{1}{2} \Big( \sigma_{++}^{++} + \sigma_{++}^{--} \Big)
+ \varepsilon \sigma_{00}^{++} 
- \varepsilon \cos(2\phi)\, \re \sigma_{+-}^{++}
- \sqrt{\varepsilon (1+\varepsilon)}\,
  \cos\phi\, \re (\sigma_{+0}^{++} + \sigma_{+0}^{--})
\phantom{\Bigg[ \Bigg] }
\nonumber \\
 && {}
- P_\ell\, \sqrt{\varepsilon (1-\varepsilon)}\, 
           \sin\phi\, \im (\sigma_{+0}^{++} + \sigma_{+0}^{--})
\phantom{\Bigg[ \Bigg] }
\nonumber \\
 && {}- S_L\, \Bigg[ 
  \varepsilon \sin(2\phi)\, \im \sigma_{+-}^{++}
+ \sqrt{\varepsilon (1+\varepsilon)}\,
  \sin\phi\, \im (\sigma_{+0}^{++} - \sigma_{+0}^{--})
\Bigg]
\nonumber \\
 && {}+ S_L P_\ell\, \Bigg[ \,
  \sqrt{1-\varepsilon^2}\; \frac{1}{2} 
  \Big( \sigma_{++}^{++} - \sigma_{++}^{--} \Big)
- \sqrt{\varepsilon (1-\varepsilon)}\,
  \cos\phi\, \re (\sigma_{+0}^{++} - \sigma_{+0}^{--})
\Bigg]
\nonumber \\
 && {}- S_T\, \Bigg[
  \sin(\phi-\phi_S)\,
  \im (\sigma_{++}^{+-} + \varepsilon \sigma_{00}^{+-})
+ \frac{\varepsilon}{2} \sin(\phi+\phi_S)\, \im \sigma_{+-}^{+-}
+ \frac{\varepsilon}{2} \sin(3\phi-\phi_S)\, \im \sigma_{+-}^{-+}
\nonumber \\
 && \hspace{2em} {}
+ \sqrt{\varepsilon (1+\varepsilon)}\, 
  \sin\phi_S\, \im \sigma_{+0}^{+-}
+ \sqrt{\varepsilon (1+\varepsilon)}\, 
  \sin(2\phi-\phi_S)\,  \im \sigma_{+0}^{-+}
\Bigg]
\nonumber \\
 && {}+ S_T P_\ell\, \Bigg[
  \sqrt{1-\varepsilon^2}\, \cos(\phi-\phi_S)\, \re \sigma_{++}^{+-}
\nonumber \\
 && \hspace{3em} {}
- \sqrt{\varepsilon (1-\varepsilon)}\, 
  \cos\phi_S\, \re \sigma_{+0}^{+-}
- \sqrt{\varepsilon (1-\varepsilon)}\, 
  \cos(2\phi-\phi_S)\,  \re \sigma_{+0}^{-+}
\Bigg] .
\end{eqnarray}
For the sake of legibility we have labeled the target spin states by
$\pm$ instead of $\pm \frac{1}{2}$.  In the following we will also use
the common notation
\begin{equation}
  \label{sigmaLT-def}
\sigma_T = \half (\sigma_{++}^{++} + \sigma_{++}^{--}) ,
\qquad\qquad
\sigma_L = \sigma_{00}^{++}
\end{equation}
for the transverse and longitudinal $\gamma^* p$ cross sections.  The
dependence of the $\ell p$ cross section on $\varepsilon$ and on the
angles $\phi$ and $\phi_S$ (or $\psi$ as explained in
Sect.~\ref{sec:transv}) is fully explicit in (\ref{Xsection-master}).

Relations analogous to (\ref{sigma-mn-ij-props}) and
(\ref{Xsection-master}) hold for cross sections and interference terms
that are differential in $M_X^2$ and $t$, or equivalently in $M_X^2$
and $\tvec{P}_{hT}^2$, where $\tvec{P}_{hT}^{\phantom{2}}$ is the
transverse component of the hadron momentum with respect to the
virtual photon momentum (see Fig.~\ref{fig:coordinates}).  Let us
analyze the behavior of the different interference terms in the region
of small $\tvec{P}_{hT}$.  To this end we go to the $\gamma^* p$
center of mass and consider the amplitudes for $\gamma^* p \to h X$ as
a function of the scattering angle $\Theta$ between $h$ and
$\gamma^*$.  For semi-inclusive processes, we can choose the set of
states $X$ to be summed over in the cross section such that the system
$X$ has definite total spin $j_X$ and definite spin projection $m_X$
along its momentum.  For exclusive processes we simply choose helicity
states of the single hadron $X$.  Also taking states with definite
helicity $m_h$ of the hadron $h$, we can perform a partial-wave
decomposition of the $\gamma^* p$ scattering amplitude (see
e.g.~\cite{Martin:1970aa}):
\begin{equation}
  \label{partial-wave}
\mathcal{A}_m^i(j_X,m_X,m_h; \Theta) = \sum_{J} 
  a_{m}^{i}(j_X,m_X,m_h; J)\; d^J_{i-m,\, m_X-m_h}(\Theta) .
\end{equation}
For $\Theta\to 0$ the rotation functions follow the behavior
$d^J_{\mu,\,\mu'}(\Theta) \sim \Theta^{|\mu-\mu'|}$.  In the product
$( \mathcal{A}^i_m )^* \mathcal{A}^j_n$ we thus have a sum over terms
which behave like $\Theta$ to the power $|i-m-m_X+m_h| + |j-n-m_X+m_h|
\ge |i-m-j+n|$.  Since $\Theta \sim |\tvec{P}_{hT}|$ for small
$\Theta$, we finally obtain a power behavior like
\begin{equation}
  \label{small-PT}
\frac{d\sigma^{ij}_{mn}}{d \tvec{P}_{hT}^2} \sim
  |\tvec{P}_{hT}|^{|m-n-i+j|} 
\qquad\qquad \mbox{for~} \tvec{P}_{hT} \to 0 ,
\end{equation}
or like a higher power of $|\tvec{P}_{hT}|$.  Applying this to our
cross section formula (\ref{Xsection-master}) we find the simple rule
that terms coming with an angular dependence $\cos(M \phi + N \phi_S)$
or $\sin(M \phi + N \phi_S)$ behave like $d\sigma_{mn}^{ij}/
(d\tvec{P}_{hT}^2) \sim |\tvec{P}_{hT}|^M$ or like a higher power,
where $M=0,1,2,3$ and $N=-1,0,1$.

Using the transformations (\ref{long-rel}) and (\ref{transv-rel}) we
obtain from (\ref{Xsection-master}) the cross sections for definite
target polarization with respect to the lepton beam,
\begin{eqnarray}
  \label{Xsection-long}
\lefteqn{
\Bigg[ \frac{\alpha_{\rm em}}{4\pi^2}\, \frac{y^2}{1-\varepsilon}\,
       \frac{1-x_B}{x_B}\, \frac{1}{Q^2} \Bigg]^{-1}
\frac{d\sigma}{dx_B\, dQ^2\, d\phi} \; \Bigg|_{P_T=0} 
  = \mbox{terms independent of $P_L$} 
}
\nonumber \\
 &-& P_L\, \Bigg[ \sin\phi\, \Big( 
   \cos\theta \sqrt{\varepsilon (1+\varepsilon)} \,
   \im (\sigma_{+0}^{++} - \sigma_{+0}^{--}) 
 - \sin\theta\, \im (\sigma_{++}^{+-} + \varepsilon \sigma_{00}^{+-})
 - \sin\theta\, \frac{\varepsilon}{2}\, \im \sigma_{+-}^{+-} \Big)
\nonumber \\[-0.1em]
 && \hspace{0.8em} {}+ \sin(2\phi) \, \Big(
   \cos\theta\, \varepsilon\, \im \sigma_{+-}^{++} 
 - \sin\theta\, \sqrt{\varepsilon (1+\varepsilon)} \, 
   \im \sigma_{+0}^{-+} \Big)
\nonumber \\
 && \hspace{0.8em} {}- \sin(3\phi) \, 
   \sin\theta\, \frac{\varepsilon}{2}\, \im \sigma_{+-}^{-+}
\Bigg] 
\nonumber \\
 &+& P_L P_\ell\, \Bigg[
   \cos\theta \sqrt{1-\varepsilon^2}\; \frac{1}{2} 
   ( \sigma_{++}^{++} - \sigma_{++}^{--} )
 + \sin\theta \sqrt{\varepsilon (1-\varepsilon)}\, 
   \re \sigma_{+0}^{+-}
\nonumber \\[-0.2em]
 && \hspace{2em} {}- \cos\phi\, \Big(
   \cos\theta \sqrt{\varepsilon (1-\varepsilon)}\, 
   \re (\sigma_{+0}^{++} - \sigma_{+0}^{--})
 + \sin\theta \sqrt{1-\varepsilon^2}\,
   \re \sigma_{++}^{+-} \Big)
\nonumber \\
 && \hspace{2em} {}+ \cos(2\phi)\, \sin\theta 
   \sqrt{\varepsilon (1-\varepsilon)}\, \re \sigma_{+0}^{-+}
\Bigg] 
\end{eqnarray}
for longitudinal and
\begin{eqnarray}
  \label{Xsection-transv}
\lefteqn{
\Bigg[ \frac{\cos\theta}{1 
     - \sin^2\!\theta\, \sin^2\!\phi_S} \Bigg]^{-1} \,
\Bigg[ \frac{\alpha_{\rm em}}{8\pi^3}\, \frac{y^2}{1-\varepsilon}\,
       \frac{1-x_B}{x_B}\, \frac{1}{Q^2} \Bigg]^{-1}
\frac{d\sigma}{dx_B\, dQ^2\, d\phi\, d\phi_S} \; \Bigg|_{P_L=0} 
}
\nonumber \\
 &=& \mbox{terms independent of $P_T$} 
     \phantom{\Bigg[ \Bigg]}
\nonumber \\
 &-& \frac{P_T}{\sqrt{1-\sin^2\!\theta \, \sin^2\!\phi_S}}\, \Bigg[
   \sin\phi_S\, \cos\theta
   \sqrt{\varepsilon (1+\varepsilon)}\,  \im \sigma_{+0}^{+-} 
   \nonumber \\[-0.8em]
 && \hspace{8.1em} {}+ \sin(\phi-\phi_S)\, \Big(
   \cos\theta\, \im (\sigma_{++}^{+-} + \varepsilon \sigma_{00}^{+-})
 + \frac{1}{2}\sin\theta \sqrt{\varepsilon (1+\varepsilon)}\,
   \im (\sigma_{+0}^{++} - \sigma_{+0}^{--}) \Big)
\nonumber \\
 && \hspace{8.1em} {}+ \sin(\phi+\phi_S)\, \Big(
   \cos\theta\, \frac{\varepsilon}{2} \, \im \sigma_{+-}^{+-}
 + \frac{1}{2}\sin\theta \sqrt{\varepsilon (1+\varepsilon)}\,
   \im (\sigma_{+0}^{++} - \sigma_{+0}^{--}) \Big)
\nonumber \\
 && \hspace{8.1em} {}+ \sin(2\phi-\phi_S)\, \Big(
   \cos\theta \sqrt{\varepsilon (1+\varepsilon)}\, 
   \im \sigma_{+0}^{-+}
 + \frac{1}{2}\sin\theta\, \varepsilon\, \im \sigma_{+-}^{++} \Big)
\nonumber \\
 && \hspace{8.1em} {}+ \sin(2\phi+\phi_S)\, \frac{1}{2}\sin\theta\, 
   \varepsilon\, \im \sigma_{+-}^{++}
\nonumber \\[-0.3em]
 && \hspace{8.1em} {}+ \sin(3\phi-\phi_S)\, \cos\theta\,
   \frac{\varepsilon}{2}\, \im \sigma_{+-}^{-+} \,
\Bigg]
\nonumber \\
 &-& \frac{P_T P_\ell}{\sqrt{1-\sin^2\!\theta \, \sin^2\!\phi_S}}\,
   \Bigg[ \cos\phi_S\, \Big(
   \cos\theta \sqrt{\varepsilon (1-\varepsilon)}\, 
   \re \sigma_{+0}^{+-}
 - \sin\theta \sqrt{1-\varepsilon^2}\; \frac{1}{2} 
   ( \sigma_{++}^{++} - \sigma_{++}^{--} ) \Big)
\nonumber \\[-0.8em]
 && \hspace{8.1em} {}-\cos(\phi-\phi_S)\, \Big(
   \cos\theta \sqrt{1-\varepsilon^2}\, \re \sigma_{++}^{+-}
 - \frac{1}{2}\sin\theta \sqrt{\varepsilon (1-\varepsilon)}\,
   \re (\sigma_{+0}^{++} - \sigma_{+0}^{--}) \Big)
\nonumber \\
 && \hspace{8.1em} {}+ \cos(\phi+\phi_S)\, \frac{1}{2}\sin\theta 
   \sqrt{\varepsilon (1-\varepsilon)}\,
   \re (\sigma_{+0}^{++} - \sigma_{+0}^{--})
\nonumber \\[-0.3em]
 && \hspace{8.1em} {}+ \cos(2\phi-\phi_S)\, \cos\theta\,
   \sqrt{\varepsilon (1-\varepsilon)}\, \re \sigma_{+0}^{-+} \,
\Bigg] 
\end{eqnarray}
for transverse polarization.  The terms independent of $P_L$ and $P_T$
are those given in the first two lines on the right-hand side of
(\ref{Xsection-master}).  Although the expressions for the
experimentally accessible cross sections (\ref{Xsection-long}) and
(\ref{Xsection-transv}) are a little lengthy, they have a clear
structure.  Using the relations $\cos\phi_S\, \sin(n \phi) =
\frac{1}{2} [\, \sin(n\phi+\phi_S) + \sin(n\phi-\phi_S) \,]$ and
$\cos\phi_S\, \cos(n \phi) = \frac{1}{2} [\, \cos(n\phi+\phi_S) +
\cos(n\phi-\phi_S) \,]$, we have written the cross sections such that
the terms in each line can be experimentally separated by measuring
the dependence on $\phi$ and (with transverse target polarization) on
$\phi_S$.  Different terms $\sigma_{mn}^{ij}(x_B,Q^2)$ multiplying the
same function of $\phi$ and $\phi_S$ can be separated by the
Rosenbluth technique, measuring at several $\ell p$ collision energies
to get several values of $\varepsilon$ at the same $x_B$ and $Q^2$.  A
different possibility is to combine data with transverse and
longitudinal target polarization.  Here one can analyze a limited
number of terms at a time:
\begin{enumerate}
\item The three combinations $\im (\sigma_{+0}^{++} -
  \sigma_{+0}^{--})$, $\im (\sigma_{++}^{+-} + \varepsilon
  \sigma_{00}^{+-})$ and $\im \sigma_{+-}^{+-}$ can be separated by
  combined analysis of the $\sin\phi$ term in the longitudinal cross
  section (\ref{Xsection-long}) and the $\sin(\phi-\phi_S)$ and
  $\sin(\phi+\phi_S)$ terms in the transverse cross section
  (\ref{Xsection-transv}).  Further separation of $\im
  \sigma_{++}^{+-}$ and $\im \sigma_{00}^{+-}$ is only possible with
  the Rosenbluth method, as is the separation of the cross sections
  $\sigma_T$ and $\sigma_L$ in $\sigma_T + \varepsilon \sigma_L$.  The
  terms just discussed are of particular physical interest, and we
  will come back to them in Sect.~\ref{sec:sidis}, \ref{sec:mesons}
  and \ref{sec:positive}.
\item The cross section difference $\sigma_{++}^{++} -
  \sigma_{++}^{--}$ and the interference term $\re \sigma_{+0}^{+-}$
  can be obtained by combined analysis of the $\phi$ independent terms
  in (\ref{Xsection-long}) and (\ref{Xsection-transv}), i.e.\ by
  integrating over $\phi$ and forming double spin asymmetries for
  polarized beam and target.  In this case, the transformation between
  asymmetries for target polarization with respect to the beam or to
  the virtual photon direction is well known from the measurement of
  the structure functions $g_1$ and $g_2$ in inclusive DIS.  We give
  the relation between the notation usually employed in the literature
  and ours in App.~\ref{app:usual-dis}.
\item $\im \sigma_{+-}^{++}$ and $\im \sigma_{+0}^{-+}$ can be
  separated by measuring both the $\sin(2\phi)$ term in
  (\ref{Xsection-long}) and the $\sin(2\phi-\phi_S)$ term in
  (\ref{Xsection-transv}).  $\im \sigma_{+-}^{++}$ can also directly
  be obtained from the $\sin(2\phi+\phi_S)$ term for transverse target
  polarization, where it is however suppressed by $\sin\theta$.  The
  situation is analogous for the $\cos\phi$ term in
  (\ref{Xsection-long}) and the $\cos(\phi-\phi_S)$ and
  $\cos(\phi+\phi_S)$ terms in (\ref{Xsection-transv}).
\item $\im \sigma_{+-}^{-+}$ can either be extracted from the
  $\sin(3\phi-\phi_S)$ term in the transverse cross section, or from
  the $\sin(3\phi)$ term in the longitudinal one, where it is however
  suppressed by $\sin\theta$.  An analogous statement holds for the
  $\cos(2\phi-\phi_S)$ term in (\ref{Xsection-transv}) and the
  $\cos(2\phi)$ term in (\ref{Xsection-long}).  Finally, the
  interference term $\im \sigma_{+0}^{+-}$ only appears in the
  transverse cross section (\ref{Xsection-transv}).
\end{enumerate}
To conclude this section we take a closer look on azimuthal moments
for transverse target polarization, which are given by
\begin{equation}
  \label{azi-moment}
\langle w(\phi,\phi_S)\, \rangle_{UT}^{\ell} = 
  \frac{\int d\phi\, d\phi_S\, w(\phi,\phi_S)\,
        [S(\phi,\phi_S) - S(\phi,\phi_S+\pi)] 
        \rule[-0.5em]{0pt}{1em}}{
        \int d\phi\, d\phi_S\,
        [S(\phi,\phi_S) + S(\phi,\phi_S+\pi)]
        \rule{0pt}{1.1em}} \;
  \Bigg|_{P_T=1, P_L=0}
\end{equation}
and similarly for the double spin asymmetry with polarized beam and
target.  For brevity we have written $S(\phi,\phi_S) = d\sigma
/(dx_B\, dQ^2\, d\phi\, d\phi_S)$ and not displayed the dependence on
$x_B$, $y$ and $Q^2$.  Without the $\phi_S$ dependence introduced by
the global factor
\begin{equation}
f(\sin^2\!\phi_S) =
  \frac{\cos\theta}{(1 - \sin^2\!\theta\, \sin^2\!\phi_S)^{3/2}
  \rule{0pt}{1.1em}}
\end{equation}
in the $P_T$ dependent part of $S(\phi,\phi_S)$, the moment of
$w(\phi,\phi_S) = 2 \sin(m\phi + \phi_S)$ would directly project out
the $\sin(m\phi + \phi_S)$ term in the transverse cross section
(\ref{Xsection-transv}), where $m= 0, \pm 1, \pm 2, -3$.  Taking the
effect of this term into account is straightforward, given that
\begin{eqnarray}
\lefteqn{
\frac{1}{2\pi^2} 
\int_0^{2\pi} d\phi \int_0^{2\pi} d\phi_S\, f(\sin^2\!\phi_S)
     \sin(n\phi+\phi_S)\, \sin(m\phi+\phi_S)
} \\
 &=&  \renewcommand{\arraystretch}{2.2}
\left\{
\begin{array}{lll}
{\displaystyle \frac{1}{\pi} \int_0^{2\pi}} d\phi_S\, 
    f(\sin^2\!\phi_S) \, \sin^2\!\phi_S 
  & = 1 + \frac{5}{8}\sin^2\!\theta + O(\sin^4\!\theta)
  & \mbox{~for~} n=m=0 , \\
{\displaystyle \frac{1}{2\pi} 
    \int_0^{2\pi}} d\phi_S\, f(\sin^2\!\phi_S)
  & = 1 + \frac{1}{4}\sin^2\!\theta + O(\sin^4\!\theta)
  & \mbox{~for~} n=m \neq 0 , \\
{\displaystyle \frac{1}{2\pi} \int_0^{2\pi}} d\phi_S\,
    f(\sin^2\!\phi_S) \, (2\sin^2\!\phi_S - 1)
  & = \frac{3}{8}\sin^2\!\theta + O(\sin^4\!\theta)
  & \mbox{~for~} n=-m \neq 0 ,
\end{array}
\right.
\nonumber \\[0.8em]
\lefteqn{
\frac{1}{2\pi^2} 
\int_0^{2\pi} d\phi \int_0^{2\pi} d\phi_S\, f(\sin^2\!\phi_S)
     \cos(n\phi+\phi_S)\, \cos(m\phi+\phi_S)
}
\nonumber \\
 &=&  \renewcommand{\arraystretch}{2.2}
\left\{
\begin{array}{lll}
{\displaystyle \frac{1}{\pi} \int_0^{2\pi}} d\phi_S\, 
    f(\sin^2\!\phi_S) \, \cos^2\phi_S 
  & = 1 - \frac{1}{8}\sin^2\!\theta - O(\sin^4\!\theta)
  & \mbox{~for~} n=m=0 , \\
{\displaystyle \frac{1}{2\pi} \int_0^{2\pi}} d\phi_S\, f(\sin^2\!\phi_S)
  & = 1 + \frac{1}{4}\sin^2\!\theta + O(\sin^4\!\theta)
  & \mbox{~for~} n=m \neq 0 , \\
{\displaystyle \frac{1}{2\pi} \int_0^{2\pi}} d\phi_S\,
    f(\sin^2\!\phi_S) \, (2\cos^2\phi_S - 1)
  & = -\frac{3}{8}\sin^2\!\theta - O(\sin^4\!\theta)
  & \mbox{~for~} n=-m \neq 0 ,
\end{array}
\right.
\nonumber
\end{eqnarray}
where for all other combinations of $m$ and $n$ the integrals are
zero.  For simplicity we have Taylor expanded the exact expressions,
which are given by elliptic integrals.  We see in particular that the
moment $\langle\, 2\sin(m \phi + \phi_S)\, \rangle_{UT}^{\ell}$
projects out not only the $\sin(m \phi + \phi_S)$ term in the cross
section but has an admixture from the $\sin(m \phi - \phi_S)$ term,
and vice versa.  This admixture comes with a prefactor $\frac{3}{8}
\sin^2\!\theta$ and therefore is typically small in deep inelastic
kinematics.  If high precision is required, one can readily invert the
linear relation between the moments $\langle\, 2\sin(m \phi +
\phi_S)\, \rangle_{UT}^{\ell}$ and $\langle\, 2\sin(m \phi - \phi_S)\,
\rangle_{UT}^{\ell}$ and the coefficients of $\sin(m \phi + \phi_S)$
and $\sin(m \phi - \phi_S)$ in the cross section.  Alternatively, one
can avoid this mixing effect by including a factor $1/
f(\sin^2\!\phi_S)$ in the weight functions $w(\phi,\phi_S)$, which
then also depend on $\theta$.

We emphasize that the results in this section do not depend on our
choice of coordinate systems.  They depend on the angles $\phi$ and
$\phi_S$ and on the phase conventions for spin states as specified in
(\ref{two-spinors}) and (\ref{photon-pol}), which can be defined
independently of a reference frame and coordinate system (see
also~\cite{Bacchetta:2004jz}).  We have used the different systems
$C$, $C'$ and $C''$ of Fig.~\ref{fig:coordinates} in order to have
simple expressions in intermediate steps of our derivation.


\section{Hadron pair production}
\label{sec:dihadron}

In a number of physically interesting cases one considers processes
\begin{equation}
  \label{two-proc}
\ell(l) + p(P) \to \ell(l') + h_1(P_1) + h_2(P_2) + X(P')
\end{equation}
with two hadrons $h_1$ and $h_2$ instead of a single hadron as in
(\ref{gen-proc}).  Examples are semi-inclusive or exclusive production
of $\pi^+\pi^-$ pairs, either from the decay of a $\rho^0$ or from the
continuum.  Let us write $P_h=P_1+P_2$ for the total momentum of the
hadron pair.  To describe the kinematics of (\ref{two-proc}) we need
three more variables in addition to the case of a single hadron $h$
(where one can e.g.\ choose $x_B$, $y$, $Q^2$, $\phi_S$, $\phi$,
$M_X^2$ and $\tvec{P}_{hT}^2$).  One additional variable is the
squared invariant mass $M_h^2 = P_h^2$ of $h_1$ and $h_2$, and the two
others can be chosen as the polar and azimuthal angles $\vartheta$,
$\varphi$ of the hadron $h_1$ in the rest frame of the pair, defined
in a coordinate system with the $z$ axis pointing opposite to
$\tvec{P}'$ and the $x$ axis taken such that $\tvec{P}$ lies in the
$x$-$z$ plane and has a positive $x$ component.\footnote{See for
instance Fig.~6 in \protect\cite{Ackerstaff:2000bz} or Fig.~5 in
\protect\cite{Berger:2001xd}, where these angles are denoted by
$(\theta,\phi)$ or $(\theta,\varphi)$, respectively.}
The invariant mass $M_h$ is invariant under a parity transformation,
and so is the polar angle $\vartheta$ (which can be expressed through
scalar products of four-vectors).  As a consequence, the relations
(\ref{sigma-mn-ij-props}) also hold for the differential cross
sections and interference terms $d\sigma_{mn}^{ij} /(dM_h^2\,
d\hspace{-1pt}\cos\vartheta)$, and our cross section formulae
(\ref{Xsection-master}), (\ref{Xsection-long}) and
(\ref{Xsection-transv}) can be made differential in $M_h^2$ and in
$\cos\vartheta$.  This is for instance important for the analysis of
exclusive pion pair production, which we will briefly discuss in
Sect.~\ref{sec:mesons}.  Note that our results do not generalize so
easily to the dependence on the azimuthal angle $\varphi$.  Since
$\varphi$ is not invariant under a parity transformation, the
relations (\ref{sigma-mn-ij-props}) no longer hold when the $\varphi$
dependence is included.  General analyses of the cross section
structure for this case can be found in \cite{Schilling:1973ag} for an
unpolarized and in \cite{Fraas:1974aa} for a polarized target.

A different generalization of the results in Sect.~\ref{sec:ep2gp} is
relevant for the analysis of semi-inclusive hadron pair production in
the framework of dihadron fragmentation functions.  This offers a way
to measure the transversity distribution of quarks in the proton, see
\cite{Bacchetta:2004kn} for a discussion and references.  In this case
the $\ell p$ cross section is required as a function not of the angle
$\phi$ between the lepton plane and the plane spanned by $\tvec{q}$
and $\tvec{P}_h$, but of the angle $\phi_R$ between the lepton plane
and the plane spanned by $\tvec{q}$ and the relative momentum
$\tvec{R}= \frac{1}{2} (\tvec{P}_1 - \tvec{P}_2)$ (with all momenta
taken in the target rest frame).  Our derivation in
Sect.~\ref{sec:ep2gp} used $\gamma^* p$ cross sections and
interference terms for polarizations defined with respect to the
$\tvec{q}$--$\tvec{P}_h$ plane, with the crucial point that this
definition only referred to the kinematics of the $\gamma^* p$
subprocess.  It is straightforward to repeat the derivation for
polarizations defined with respect to the $\tvec{q}$--$\tvec{R}$
plane, and the result will be the analogs of the cross section
formulae (\ref{Xsection-master}), (\ref{Xsection-long}) and
(\ref{Xsection-transv}) with $\phi$ replaced by $\phi_R$, and with
$\gamma^* p$ cross sections and interference terms referring to
different polarization states than in Sect.~\ref{sec:ep2gp}.  The
cross sections $\sigma_{mm}^{ii}$ and the interference term
$\sigma_{+0}^{+-}$ are actually the same in both cases, since they
appear without a $\phi$ or $\phi_R$ dependence, but all other
interference terms will in general depend on the choice of
polarization states.


\section{Semi-inclusive deep inelastic scattering}
\label{sec:sidis}

\begin{figure}
\begin{center}
\leavevmode
\includegraphics[width=0.85\textwidth]{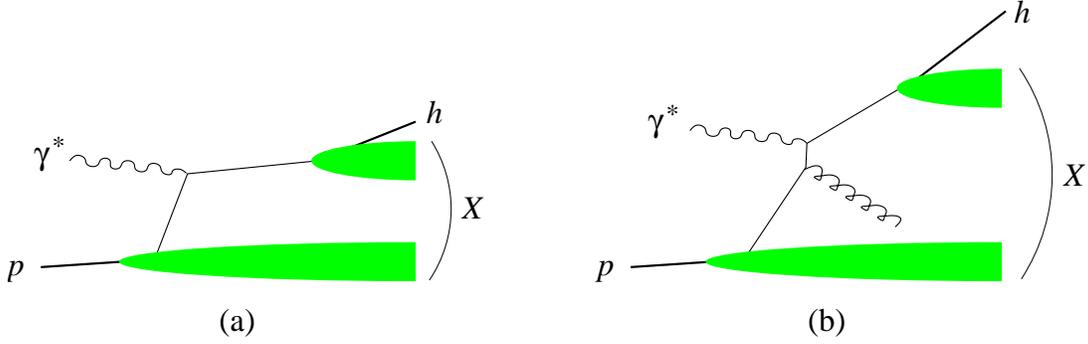}
\end{center}
\caption{\label{fig:sidis} Semi-inclusive hadron production $\gamma^*
  p\to h X$ at large $Q^2$.  (a) Born level graph.  (b) A
  next-to-leading order graph where the hadron $h$ has transverse
  momentum of order $Q$.}
\end{figure}

Let us now take a closer look at semi-inclusive hadron production.  It
is customary to trade the variable $M_X^2$ for $z = (P_h \cdot P)
/(q\cdot P)$ in this case.  In the kinematical limit of large $Q^2$ at
given $x_B$, $z$ and $\tvec{P}_h$, the cross section factorizes into a
hard-scattering subprocess multiplied with parton densities and
fragmentation functions that explicitly depend on the transverse
parton momentum.  The corresponding Born level expressions have been
calculated in
\cite{Mulders:1995dh,Boer:1997nt,Boer:1999uu,Bacchetta:2004zf} at
leading and first subleading order in $1/Q$.  We will remark on
$\alpha_s$ effects at the end of this section.  The Born level results
show a simple pattern:
\begin{enumerate}
\item At leading order in $1/Q$ we have cross sections and
  interference terms $\sigma_{++}^{ij}$ and $\sigma_{+-}^{ij}$ that
  involve only transverse photon polarization.  They are expressed in
  terms of twist-two parton densities and twist-two fragmentation
  functions.  These twist-two functions have a simple
  probabilistic interpretation in the parton model. see
  e.g.~\cite{Boglione:1999pz,Barone:2001sp}.
\item The interference terms $\sigma_{+0}^{ij}$ between a transverse
  and a longitudinal photon are suppressed by one power of $1/Q$.
  They involve a twist-two parton density times a twist-three
  fragmentation function or vice versa.
\item The longitudinal cross section $\sigma_{00}^{++}$ and the
  interference term $\sigma_{00}^{+-}$ do not appear in the result.
  At Born level they must hence be suppressed by $1/Q^2$.
\end{enumerate}
For brevity we will in the following refer to the cross sections and
interference terms of point 1 as ``twist-two'' and to those of point 2
as ``twist-three'' quantities.  The finding in point 1 has a simple
physical reason.  To leading accuracy in $1/Q$ the transverse momentum
and the virtuality of the incoming and outgoing parton is set to zero
when evaluating the hard scattering, which at Born level is just the
scattering of a quark or antiquark on a virtual photon, see
Fig.~\ref{fig:sidis}a.  In the Breit frame one readily sees that
conservation of the fermion helicity requires the photon to have
transverse polarization.  This is the well-known mechanism responsible
for the Callan-Gross relation in inclusive DIS.

For definiteness let us express the leading-twist results from
\cite{Boer:1999uu} in terms of $\gamma^* p$ cross sections and
interference terms.  Using the abbreviation
\begin{equation}
\Gamma = \frac{4\pi^3 \alpha_{\rm em}}{Q^2}\, \frac{x_B}{1-x_B}
\end{equation}
we can write
\begin{eqnarray}
  \label{amsterdam-results}
\frac{1}{2}\, \Bigg[
   \frac{d\sigma_{++}^{++}}{dz\, d\tvec{P}_{hT}^2} 
 + \frac{d\sigma_{++}^{--}}{dz\, d\tvec{P}_{hT}^2} \Bigg]
 &=& \Gamma\, 
  \mathcal{F}\Big[ f_1 D_1 \Big] ,
\nonumber \\
\frac{1}{2}\, \Bigg[
   \frac{d\sigma_{++}^{++}}{dz\, d\tvec{P}_{hT}^2} 
 - \frac{d\sigma_{++}^{--}}{dz\, d\tvec{P}_{hT}^2} \Bigg]
 &=& \Gamma\, 
  \mathcal{F}\Big[ g_1 D_1 \Big] ,
\nonumber \\
\re \frac{d\sigma_{+-}^{++}}{dz\, d\tvec{P}_{hT}^2} 
 &=& \Gamma\, 
  \mathcal{F}\Bigg[\, \frac{|\tvec{p}_T|\, |\tvec{k}_T|
   \cos(\varphi_p + \varphi_k)}{M_p\, M_h}\;
   h_1^\perp H_1^\perp \,\Bigg] ,
\nonumber \\
\im \frac{d\sigma_{+-}^{++}}{dz\, d\tvec{P}_{hT}^2} 
 &=& \Gamma\, 
  \mathcal{F}\Bigg[\, \frac{|\tvec{p}_T|\, |\tvec{k}_T|
   \cos(\varphi_p + \varphi_k)}{M_p\, M_h}\;
   h_{1L}^\perp H_1^\perp \,\Bigg] ,
\nonumber \\
\re \frac{d\sigma_{++}^{+-}}{dz\, d\tvec{P}_{hT}^2} 
 &=& \Gamma\, 
  \mathcal{F}\Bigg[\, \frac{|\tvec{p}_T| \cos\varphi_p}{M_p}\;
   g_{1T}  D_1 \,\Bigg] ,
\nonumber \\
\im \frac{d\sigma_{++}^{+-}}{dz\, d\tvec{P}_{hT}^2} 
 &=& \Gamma\, 
  \mathcal{F}\Bigg[\, \frac{|\tvec{p}_T| \cos\varphi_p}{M_p}\;
   f_{1T}^\perp  D_1^{\phantom{\perp}} \,\Bigg] ,
\nonumber \\
\im \frac{d\sigma_{+-}^{+-}}{dz\, d\tvec{P}_{hT}^2} 
 &=& \Gamma\, 
  \mathcal{F}\Bigg[\, \frac{|\tvec{k}_T| \cos\varphi_k}{M_h}\;
   2 h_1^{\phantom{\perp}\!\!\!} H_1^\perp \,\Bigg] ,
\nonumber \\
\im \frac{d\sigma_{+-}^{-+}}{dz\, d\tvec{P}_{hT}^2} 
 &=& \Gamma\, 
  \mathcal{F}\Bigg[\, \frac{|\tvec{p}_T|^2 |\tvec{k}_T|
   \cos(2\varphi_p + \varphi_k)}{M_p^2\, M_h}\;
   h_{1T}^\perp H_1^\perp \,\Bigg] ,
\hspace{4em}
\end{eqnarray}
with convolution integrals given by
\begin{equation}
  \label{amsterdam-convolutions}
\mathcal{F}\Big[ w f D \Big] = \sum_{a=q, \bar{q}} e_a^2\, 
  \int d^2\tvec{p}_T\, d^2\tvec{k}_T^{\phantom{*}}\, 
  \delta^{(2)}\Big( \tvec{p}_T^{\phantom{*}} 
                  - \tvec{k}_T^{\phantom{*}} 
                  - \tvec{P}_{hT}^{\phantom{*}} /z \Big)\;
  w(\tvec{p}_T, \tvec{k}_T^{\phantom{*}})\,
  f^a(x_B,\tvec{p}_T^2)\; D^a(z, z^2 \tvec{k}_T^2) ,
\end{equation}
where $f$ represents a parton density, $D$ a fragmentation function,
and $w$ an additional weight function.  To write the weight functions
in a compact way we have used angles $\varphi_{p} = \angle
(\tvec{p}_T^{\phantom{*}}, \tvec{P}_{hT}^{\phantom{*}})$ and
$\varphi_{k} = \angle (\tvec{k}_T, \tvec{P}_{hT})$ in the transverse
plane.  The quark or antiquark densities (in lowercase symbols) depend
on $x_B$ and on the transverse momentum $\tvec{p}_T$ of the parton
relative to the proton.  The fragmentation functions (in uppercase
symbols) depend on $z$ and on the transverse momentum $\tvec{k}_T$ of
the parton relative to the hadron $h$ (or the transverse momentum $-z
\tvec{k}_T$ of $h$ relative to the parton).\footnote{See
\protect\cite{Mulders:1995dh} for a discussion of adequate reference
frames in this context.  We also remark that $\tvec{P}_{h\perp}$ in
the notation of \protect\cite{Mulders:1995dh} is the same as
$\tvec{P}_{hT}$ in the present paper.}
We note that some of the convolutions in (\ref{amsterdam-results})
acquire an explicit minus sign when the integrals over $\tvec{p}_T$
and $\tvec{k}_T^{\phantom{*}}$ are carried out, see e.g.\ App.~D in
\cite{Mulders:1995dh}.  As remarked in \cite{Boer:1997nt}, the
convolutions (\ref{amsterdam-convolutions}) factorize into separate
transverse momentum integrals over parton densities and over
fragmentation functions if one forms weighted cross sections $\int
d\tvec{P}_{hT}^2\, |\tvec{P}_{hT}^{\phantom{*}} /z \,|^M \,
d\sigma_{mn}^{ij} /(dz\, d\tvec{P}_{hT}^2)$ with the power $M=
|m-n-i+j|$ we encountered in (\ref{small-PT}).  We shall not discuss
all parton densities and fragmentation functions here (see
\cite{Mulders:1995dh,Boer:1997nt} for their definitions and
\cite{Boglione:1999pz} for an overview), but point out two terms of
particular interest in ongoing and planned experiments
\cite{Airapetian:2004tw,Pagano:2005jx,HallA:2003,JLab:12}.  The Sivers
function $f_{1T}^\perp$ together with the usual unpolarized
fragmentation function $D_1$ appears in $\im \sigma_{++}^{+-}$, and
the transversity distribution $h_1$ comes together with the Collins
fragmentation function $H_1^\perp$ in $\im \sigma_{+-}^{+-}$.  Many
investigations have shown these functions to reveal subtle aspects of
the dynamics and the structure of hadrons, see
e.g.~\cite{Barone:2001sp,Vogelsang:2003eb} for recent reviews.

We notice in (\ref{amsterdam-results}) that all possible cross
sections and interference terms with transverse photons are nonzero.
The results of \cite{Mulders:1995dh,Boer:1997nt,Bacchetta:2004zf} show
that all interference terms $\sigma_{+0}^{ij}$ are nonvanishing as
well.\footnote{This is in contrast to semi-inclusive hadron-pair
production in dependence of the angle $\phi_R$ discussed in
Sect.~\protect\ref{sec:dihadron}, where the calculation of
\protect\cite{Bacchetta:2003vn} gives zero entries for several
interference terms $\sigma_{++}^{ij}$, $\sigma_{+-}^{ij}$ and
$\sigma_{+0}^{ij}$.}
Taking into account the $Q^2$ behavior specified above and keeping in
mind that $\sin\theta$ is of order $1/Q$, we can now discuss the
relative size of terms which have the same dependence on $\phi$ and
$\phi_S$ in the cross sections (\ref{Xsection-long}) and
(\ref{Xsection-transv}) for definite target polarization with respect
to the beam.
\begin{enumerate}
\item For longitudinal target polarization the Sivers and Collins
  terms, $\im \sigma_{++}^{+-}$ and $\im \sigma_{+-}^{+-}$, come with
  a factor $\sin\theta$ and thus appear with the same power in $1/Q$
  as the twist-three interference term $\im(\sigma_{+0}^{++} -
  \sigma_{+0}^{--})$.  For a transversely polarized target, this
  twist-three term is multiplied with $\sin\theta$ and thus suppressed
  by $1/Q^2$ compared with the Sivers or the Collins term.
  Furthermore, the Sivers term $\im \sigma_{++}^{+-}$ always comes
  together with $\im \sigma_{00}^{+-}$, which is $1/Q^2$ suppressed
  according to our above discussion.
\item For the $\phi$ independent terms in the cross section the
  situation is reverse (and well-known from inclusive DIS).  Here it
  is for transverse target polarization that both the twist-two cross
  section difference $\sigma_{++}^{++} - \sigma_{++}^{--}$ and the
  twist-three term $\re\sigma_{+0}^{+-}$ appear with the same power in
  $1/Q$.  For a longitudinally polarized target this twist-three term
  is accompanied by $\sin\theta$ and thus down by $1/Q^2$ compared
  with $\sigma_{++}^{++} - \sigma_{++}^{--}$.
\end{enumerate}
For the terms with $\cos\phi$ and $\cos(\phi-\phi_S)$ in the polarized
cross sections (\ref{Xsection-long}) and (\ref{Xsection-transv}) the
situation is as in case 1, and for the terms with $\sin(2\phi)$ and
$\sin(2\phi-\phi_S)$ as in case 2.  

In those cases where a ``competing term'' in the cross section is
suppressed by $1/Q^2$, one may argue that it should be consistently
neglected in an analysis based on theoretical results with accuracy
only up to order $1/Q$.  After all, quantities like
$\im\sigma_{++}^{+-}$ have themselves $1/Q^2$ suppressed contributions
in addition to the leading-twist part which one would like to extract.
In general, subtracting one particular type of power-suppressed term
from an observable can improve the comparison with leading-twist
theory, but it can also make it worse, since different
power-suppressed terms may have opposite sign and partially compensate
each other.  Our case is however special.  Taking for example the
$1/Q^2$ suppressed quantities $\varepsilon \im \sigma_{00}^{+-}$ and
$\sin\theta \sqrt{\varepsilon (1+\varepsilon)}\, \im(\sigma_{+0}^{++}
- \sigma_{+0}^{--})$, which compete with $\im \sigma_{++}^{+-}$ in the
$\sin(\phi-\phi_S)$ term, we see that they come with a different
dependence on $\varepsilon$.  Since the $\sigma_{mn}^{ij}$ are
independent of this variable, these power-suppressed terms can in
general not compensate power corrections in $\im\sigma_{++}^{+-}$
itself.  This provides some theoretical motivation to try and separate
such contributions, which may be of practical relevance especially if
a twist-two term is ``accidentally'' small because the relevant parton
distributions or fragmentation functions are.

Let us finally remark on loop corrections to the Born level formulae
on which we have based our discussion so far.  At leading accuracy in
$1/Q$ these have been recently investigated in
\cite{Ji:2004wu,Collins:2004nx}.  Note that at next-to-leading order
in $\alpha_s$ there are hard-scattering graphs where two partons with
transverse momenta of order $Q$ are produced, see
Fig.~\ref{fig:sidis}b.  It was emphasized in \cite{Ji:2004wu} that
such graphs do not contribute when $\tvec{P}_{hT}$ is small compared
with $Q$ and can be generated from the transverse momentum dependence
in the parton densities and fragmentation functions, as expressed in
(\ref{amsterdam-convolutions}).  They do however contribute if one
integrates the cross section over all $\tvec{P}_{hT}$ (or takes
$\tvec{P}_{hT}$ weighted cross sections as mentioned above).  They
produce effects at leading order in $1/Q$ and can be evaluated using
standard collinear factorization, with parton densities and
fragmentation functions that are integrated over the transverse parton
momentum.  In particular, these graphs generate an order $\alpha_s$
contribution to the longitudinal cross section $\sigma_{00}^{++}$,
just as in the well-known case of inclusive DIS.  Explicit calculation
for an unpolarized target shows that they also generate a $\cos\phi$
and $\cos(2\phi)$ modulation in the cross section
\cite{Georgi:1977tv}, described by the interference terms
$\re(\sigma_{+0}^{++} + \sigma_{+0}^{--})$ and $\re\sigma_{+-}^{++}$.
The lepton polarization dependence for an unpolarized target is due to
$\im(\sigma_{+0}^{++} + \sigma_{+0}^{--})$.  Because of time reversal
invariance, this term requires an absorptive part in the amplitude and
thus appears only at order $\alpha_s^2$ in the large $\tvec{P}_{hT}$
region \cite{Hagiwara:1982cq}.


\section{Exclusive meson production}
\label{sec:mesons}

\begin{figure}
\begin{center}
\leavevmode
\includegraphics[width=0.8\textwidth]{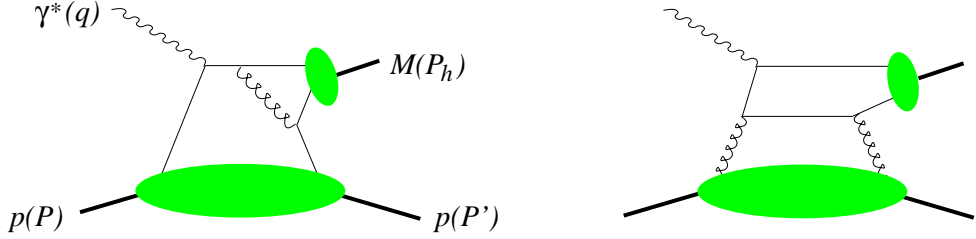}
\end{center}
\caption{\label{fig:mesons} Example graphs for exclusive production of
  a meson $M$ at large $Q^2$.  Instead of the proton there may be a
  different baryon in the final state.  The lower blobs represent
  twist-two generalized parton distributions, and the upper blobs
  stand for the twist-two distribution amplitude of the meson.}
\end{figure}

Exclusive electroproduction of light mesons such as $\ell p\to \ell
\rho^0 p$ or $\ell p\to \ell \pi^+ n$ provides opportunities to study
generalized parton distributions (GPDs), see
\cite{Goeke:2001tz,Diehl:2003ny} for recent reviews.  In the limit of
large $Q^2$ at fixed $x_B$ and $t$, the $\gamma^* p$ amplitude
factorizes into the convolution of a hard-scattering subprocess with
generalized parton distributions in the nucleon and the light-cone
distribution amplitude of the produced meson (see
Fig.~\ref{fig:mesons}).  The factorization theorem shows that the
leading transitions in the large $Q^2$ limit have both the virtual
photon and the produced meson longitudinally polarized, all other
transitions being suppressed by at least one power of $1/Q$
\cite{Collins:1996fb,Diehl:1998pd}.  This gives a hierarchy opposite
to the one we have encountered for semi-inclusive production in
Sect.~\ref{sec:sidis}:
\begin{enumerate}
\item The only leading-twist observables are the longitudinal cross
  section $\sigma_{00}^{++}$ and the interference term
  $\sigma_{00}^{+-}$.
\item Transverse-longitudinal interference terms $\sigma_{+0}^{ij}$
  are at least one power of $1/Q$ down compared with
  $\sigma_{00}^{++}$.
\item Cross sections and interference terms $\sigma_{++}^{ij}$ and
  $\sigma_{+-}^{ij}$ with transverse photon polarization are
  suppressed by at least $1/Q^2$ compared with $\sigma_{00}^{++}$.
\end{enumerate}
Using the abbreviation
\begin{equation}
\Gamma' = \frac{\alpha_{\rm em}}{Q^6}\, \frac{x_B^2}{1-x_B}
\end{equation}
the leading-twist results given in \cite{Goeke:2001tz,Diehl:2003ny}
can written as\footnote{The relation between the angle $\beta$ used in
\protect\cite{Goeke:2001tz} and the angles used here is
$\sin\beta_{\mbox{\tiny\protect\cite{Goeke:2001tz}}} = -
\sin(\phi-\phi_S)_{\mbox{\tiny here}}\,$.}
\begin{eqnarray}
  \label{meson-natural}
\frac{1}{\Gamma'}\, \frac{d\sigma_{00}^{++}}{dt}
  &=& (1-\xi^2)\, |\mathcal{H}_M|^2
   - \left( \xi^2 + \frac{t}{4M_p^2} \right) |\mathcal{E}_M|^2 
   - 2 \xi^2\, \re (\mathcal{E}_M^* \mathcal{H}_M^{\phantom{*}} ) ,
\nonumber \\
\frac{1}{\Gamma'}\; \im \frac{d\sigma_{00}^{+-}}{dt}
 &=& {}- \sqrt{1-\xi^2}\, \frac{\sqrt{t_0-t}}{M_p}\,
     \im ( \mathcal{E}_M^* \mathcal{H}_M^{\phantom{*}} )
\end{eqnarray}
for mesons with natural parity like $\rho^0$, $\rho^+$, $f_2$, and as
\begin{eqnarray}
  \label{meson-unnatural}
\frac{1}{\Gamma'}\, \frac{d\sigma_{00}^{++}}{dt}
  &=& (1-\xi^2)\, |\tilde\mathcal{H}_M|^2
   - \xi^2 \frac{t}{4M_p^2}\, |\tilde\mathcal{E}_M|^2 - 2\xi^2\, 
     \re (\tilde\mathcal{E}_M^* \tilde\mathcal{H}_M^{\phantom{*}} ) ,
  \hspace{2.8em}
\nonumber \\
\frac{1}{\Gamma'}\; \im \frac{d\sigma_{00}^{+-}}{dt}
 &=& \sqrt{1-\xi^2}\, \frac{\sqrt{t_0-t}}{M_p}\, \xi\, 
     \im ( \tilde\mathcal{E}_M^* \tilde\mathcal{H}_M^{\phantom{*}} )
\end{eqnarray}
for mesons with unnatural parity like $\pi^0$, $\pi^+$, $\eta$.  In
the kinematical factors on the right-hand side\footnote{Their
expressions for the case where outgoing baryon is not a nucleon can be
found in \protect\cite{Diehl:2003qa}.}
we have used the scaling variable $\xi$ and the smallest kinematically
allowed momentum transfer $-t_0$, given by
\begin{equation}
  \label{excl-vars}
 \xi = \frac{x_B}{2-x_B} , \qquad\qquad\qquad
-t_0 =  \frac{4\xi^2 M_p^2}{1-\xi^2} 
\end{equation}
up to relative corrections of order $x_B M_p^2 /Q^2$, $x_B\, t /Q^2$
and $M_h^2 /Q^2$.  Note that $\sqrt{t_0-t} \propto |\tvec{P}_{hT}|$,
so that the behavior of $\im d\sigma_{00}^{+-} /dt$ illustrates our
general result (\ref{small-PT}).

The quantities $\mathcal{H}_M$, $\mathcal{E}_M$,
$\tilde\mathcal{H}_M$, $\tilde\mathcal{E}_M$ are integrals over the
GPDs $H$, $E$, $\tilde{H}$, $\tilde{E}$ appropriate for the production
of the meson $M$ (given in App.~\ref{app:mesons} for $\ell p\to \ell
\rho^0 p$ and $\ell p\to \ell \pi^+ n$).  They depend on $\xi$, $t$
and $Q^2$, where the dependence on $Q^2$ is only logarithmic and
reflects the familiar scaling violations from loop corrections to the
hard-scattering kernels.  We note that for mesons with natural parity
both quark and gluon GPDs in general contribute at leading order in
$\alpha_s$, whereas for mesons with unnatural parity only quark
distributions appear at this accuracy \cite{Diehl:2003ny}.

The interest of measuring $\im d\sigma_{00}^{+-} /dt$ in addition to
$d\sigma_{00}^{++} /dt$ is immediately clear from
(\ref{meson-natural}) and (\ref{meson-unnatural}).  The combination of
these two observables provides a handle to separate the contributions
from the GPDs $H$ and $E$ or $\tilde{H}$ and $\tilde{E}$, which
describe different spin dependence.\footnote{Unfortunately, these two
observables are insufficient to uniquely determine both the size of
the convolutions $\mathcal{H}_M$ and $\mathcal{E}_M$ or
$\tilde\mathcal{H}_M$ and $\tilde\mathcal{E}_M$ and their relative
phase.}
The nucleon helicity-flip distributions $E^q$ and $E^g$ are of
particular interest because they carry information about the
contribution from the orbital angular momentum of quarks and gluons to
the total spin of the proton \cite{Ji:1996ek}.  With the $Q^2$
behavior discussed above, we find from (\ref{Xsection-transv}) that
with transverse target polarization one can obtain $\im
\sigma_{00}^{+-}$ from the $\sin(\phi-\phi_S)$ dependent term of the
$\ell p$ cross section, where it comes together with the terms $\im
\sigma_{++}^{+-}$ and $\sin\theta\, \im(\sigma_{+0}^{++} -
\sigma_{+0}^{--})$, both of which are suppressed by at least $1/Q^2$.
In the cross section (\ref{Xsection-long}) for longitudinal target
polarization, $\sin\theta\, \im\sigma_{00}^{+-}$ and
$\im(\sigma_{+0}^{++} - \sigma_{+0}^{--})$ contribute to the
$\sin\phi$ dependence with the same power of $1/Q$, together with at
least $1/Q^2$ suppressed terms $\sin\theta\, \im \sigma_{++}^{+-}$ and
$\sin\theta\, \im \sigma_{+-}^{+-}$.  We note that a nonzero effect
for this $\sin\phi$ modulation has been measured in $ep\to e\pi^+ n$
by HERMES \cite{Airapetian:2001iy}.

As discussed in the previous section, one may want to extract separate
$\gamma^* p$ cross sections and interference terms without an a priori
assumption on their relative size.  The leading-twist interference
term $\im\sigma_{00}^{+-}$ in (\ref{meson-natural}) could for instance
be ``accidentally'' small because $\mathcal{E}_M$ is much smaller than
$\mathcal{H}_M$ or because their relative phase is close to zero.
Combining data for transverse and longitudinal target polarization one
can separate the terms $\im ( \sigma_{++}^{+-} + \varepsilon
\sigma_{00}^{+-})$, $\im \sigma_{+-}^{+-}$ and $\im(\sigma_{+0}^{++} -
\sigma_{+0}^{--})$, provided that one measures both the
$\sin(\phi-\phi_S)$ and $\sin(\phi+\phi_S)$ dependence for a
transversely polarized target.  Without the Rosenbluth technique one
can however not isolate the longitudinal contribution in $\im (
\sigma_{++}^{+-} + \varepsilon \sigma_{00}^{+-})$, nor the
longitudinal part from $\sigma_T + \varepsilon \sigma_L$ in the
unpolarized cross section.

For electroproduction of vector mesons one experimentally finds that
the ratio $\sigma_L /\sigma_T$ is not very large for $Q^2$ of a few
GeV$^2$ \cite{Ackerstaff:2000bz,Adams:1997bh}, which means that the
predicted power suppression of transverse photon amplitudes is
numerically not yet very effective in this kinematics.  In addition
one finds that transitions with the same helicity for photon and meson
are clearly larger than those changing the helicity
\cite{Ackerstaff:2000bz,Adams:1997bh}, which is commonly referred to
as approximate $s$-channel helicity conservation.  The largest power
suppressed amplitudes are hence those from a transverse photon to a
transverse vector meson.  A possibility to remove this particularly
important type of power correction in an analysis is to measure the
decay angular distribution of the vector meson, say in $\rho\to
\pi^+\pi^-$.  Here we can make use of our result in
Sect.~\ref{sec:dihadron}.  If only the dependence on the polar decay
angle $\vartheta$ but not the azimuth $\varphi$ is considered, our
cross section formulae (\ref{Xsection-long}) and
(\ref{Xsection-transv}) can be made differential in $\cos\vartheta$.
Different helicities of the $\rho$ do not interfere if $\varphi$ is
integrated over, so that for all $m,n$ and $i,j$ we have
\begin{equation}
  \label{rhoTL}
\frac{d\sigma_{mn}^{ij}(\gamma^* p\to
      \pi^+\pi^- p)}{d(\cos\vartheta)} 
= \frac{3 \cos^2\!\vartheta}{2}\,
  \sigma_{mn}^{ij}(\gamma^* p\to \rho_L \hspace{1pt} p)
+ \frac{3\sin^2\!\vartheta}{4}\,
  \sigma_{mn}^{ij}(\gamma^* p\to \rho_T \hspace{1pt} p)
\end{equation}
with $\gamma^* p$ cross sections and interference terms for
longitudinal and transverse $\rho$ polarization.  Since
$\sigma_{++}^{ij}(\rho_L)$ is the product of two $s$-channel helicity
nonconserving amplitudes, it should be negligible in
$\sigma_{++}^{ij}(\rho_L) + \varepsilon \sigma_{00}^{ij}(\rho_L)$,
unless $\varepsilon$ is small.  Using the $\vartheta$ dependence in
(\ref{rhoTL}) to project out the $\rho_L$ contribution from the
$\sin(\phi-\phi_S)$ term in the cross section will hence help toward
isolating the twist-two observable $\sigma_{00}^{+-}(\rho_L)$.

We finally mention that an angular analysis analogous to (\ref{rhoTL})
can also be performed for the production of continuum $\pi^+\pi^-$
pairs, where one can measure the interference between partial waves
with different total spin of the pion pair, see
\cite{Lehmann-Dronke:2000xq,Diehl:2003ny} and
\cite{Airapetian:2004sy}.  The $\vartheta$ dependence for interference
terms $d\sigma_{mn}^{+-} /(d\cos\vartheta)$ is the same as for the
terms $d\sigma_{mn}^{++} /(d\cos\vartheta)$ accessible with an
unpolarized target.


\section{Positivity constraints}
\label{sec:positive}

In Sect.~\ref{sec:ep2gp} we introduced $\gamma^* p$ cross sections and
interference terms for specific polarization states.  The $\gamma^* p$
cross section must be positive or zero for \emph{any} polarization
state of the photon-proton system, so that $\sum_{ijmn} (c_m^i)^*\,
\sigma_{mn}^{ij}\, c_n^{\hspace{1pt}j} \ge 0$ for arbitrary complex
coefficients $c_m^i$.  This means that the matrix
\begin{equation}
  \label{bigmatrix}
M =
\renewcommand{\arraystretch}{1.4}
\left( \begin{array}{cccccc}
    \sigma_{00}^{++}       & i\, \im\sigma_{00}^{+-}
&  (\sigma_{+0}^{++})^*    &  (\sigma_{+0}^{-+})^*
&  \hspace{-1.7ex}
 -(\sigma_{+0}^{--})^*     &  (\sigma_{+0}^{+-})^* \\
 -i\, \im \sigma_{00}^{+-} &   \sigma_{00}^{++}
& (\sigma_{+0}^{+-})^*     &  (\sigma_{+0}^{--})^*
& (\sigma_{+0}^{-+})^*     &   \hspace{-1.7ex}
                             -(\sigma_{+0}^{++})^* \\
   \sigma_{+0}^{++}        &   \sigma_{+0}^{+-} 
&  \sigma_{++}^{++}        &   \sigma_{++}^{+-}
&  \sigma_{+-}^{++}        & i\, \im\sigma_{+-}^{+-} \\
   \sigma_{+0}^{-+}        &   \sigma_{+0}^{--}
& (\sigma_{++}^{+-})^*     &   \sigma_{++}^{--}
&  i\, \im\sigma_{+-}^{-+} &  (\sigma_{+-}^{++})^* \\
   \hspace{-1.7ex}
  -\sigma_{+0}^{--}        &   \sigma_{+0}^{-+}
& (\sigma_{+-}^{++})^*     &  -i\, \im\sigma_{+-}^{-+}
&  \sigma_{++}^{--}        &   \hspace{-1.7ex}
                             -(\sigma_{++}^{+-})^* \\
   \sigma_{+0}^{+-}        &   \hspace{-1.7ex}
                              -\sigma_{+0}^{++}
& -i\, \im\sigma_{+-}^{+-} &   \sigma_{+-}^{++}
&  \hspace{-1.7ex}
  -\sigma_{++}^{+-}        &   \sigma_{++}^{++}
\end{array} \right)
\end{equation}
formed by $M_{(mi) (nj)} = \sigma_{mn}^{ij}$ must be positive
semidefinite, where the rows and columns are ordered such that they
correspond to the combinations $(0,+\frac{1}{2}), (0,-\frac{1}{2}),
(+1,+\frac{1}{2}), (+1,-\frac{1}{2}), (-1,+\frac{1}{2}),
(-1,-\frac{1}{2})$ of photon and proton helicities.  In writing down
(\ref{bigmatrix}) we have used the relations (\ref{sigma-mn-ij-props})
from hermiticity and parity invariance.  We have not been able to find
closed expressions for the eigenvalues of this matrix (and if they
existed, they might be too complicated to be useful in practice).
More tractable sets of positivity bounds can be obtained if one uses
that submatrices of $M$ are also positive semidefinite.  As simple
example is the submatrix for longitudinal photons, formed from the
first and second rows and columns of $M$, whose positivity implies
\begin{equation}
  \label{long-pos}
|\, \im\sigma_{00}^{+-} | \le \sigma_{00}^{++} .
\end{equation}
The submatrix for transverse photons, formed by the third to sixth
rows and columns of $M$, has eigenvalues
\begin{eqnarray}
  \label{transv-eigenvals}
\lefteqn{2 e_{1,2} =
  \sigma_{++}^{++} - \im\sigma_{+-}^{+-}
+ \sigma_{++}^{--} + \im\sigma_{+-}^{-+}
}
\nonumber \\
&& {}\pm \sqrt{ (\sigma_{++}^{++} - \im\sigma_{+-}^{+-}
               - \sigma_{++}^{--} - \im\sigma_{+-}^{-+})^2 
   + 4 (\re\sigma_{+-}^{++} - \im\sigma_{++}^{+-})^2
   + 4 (\re\sigma_{++}^{+-} + \im\sigma_{+-}^{++})^2 } \; ,
\nonumber \\[0.4em]
\lefteqn{2 e_{3,4} =
  \sigma_{++}^{++} + \im\sigma_{+-}^{+-}
+ \sigma_{++}^{--} - \im\sigma_{+-}^{-+}
}
\nonumber \\
&& {}\pm \sqrt{ (\sigma_{++}^{++} + \im\sigma_{+-}^{+-}
               - \sigma_{++}^{--} + \im\sigma_{+-}^{-+})^2 
   + 4 (\re\sigma_{+-}^{++} + \im\sigma_{++}^{+-})^2
   + 4 (\re\sigma_{++}^{+-} - \im\sigma_{+-}^{++})^2 } \; .
\hspace{3em}
\end{eqnarray}
Note that $e_3$ and $e_4$ are obtained from $e_1$ and $e_2$ by
changing the signs of all proton helicity-flip terms
$\sigma_{mn}^{+-}$.  All four eigenvalues (\ref{transv-eigenvals})
must be nonnegative, which implies
\begin{equation}
  \label{transv-pos-1}
|\, \im \sigma_{+-}^{+-} | \le \sigma_{++}^{++} , \qquad\qquad
|\, \im \sigma_{+-}^{-+} | \le \sigma_{++}^{--} ,
\end{equation}
and
\begin{eqnarray}
  \label{transv-pos-2}
  \Big( \re\sigma_{+-}^{++} - \im\sigma_{++}^{+-} \Big)^2
+ \Big( \re\sigma_{++}^{+-} + \im\sigma_{+-}^{++} \Big)^2
&\le & \Big( \sigma_{++}^{++} - \im\sigma_{+-}^{+-} \Big) \,
       \Big( \sigma_{++}^{--} + \im\sigma_{+-}^{-+} \Big) ,
\nonumber \\
  \Big( \re\sigma_{+-}^{++} + \im\sigma_{++}^{+-} \Big)^2
+ \Big( \re\sigma_{++}^{+-} - \im\sigma_{+-}^{++} \Big)^2
&\le & \Big( \sigma_{++}^{++} + \im\sigma_{+-}^{+-} \Big) \,
       \Big( \sigma_{++}^{--} - \im\sigma_{+-}^{-+} \Big) .
\end{eqnarray}
One can easily obtain inequalities that are weaker than
(\ref{transv-pos-2}) but involve fewer interference terms, e.g.\ by
omitting one of the squared terms on the left-hand-sides.  Adding the
bounds (\ref{transv-pos-2}) one has
\begin{equation}
  \label{simpler-bound}
  ( \re\sigma_{+-}^{++} )^2 + ( \im\sigma_{++}^{+-} )^2
+ ( \re\sigma_{++}^{+-} )^2 + ( \im\sigma_{+-}^{++} )^2
\le ( \sigma_{++}^{++} )\, ( \sigma_{++}^{--} )
  - ( \im\sigma_{+-}^{+-} )\, ( \im\sigma_{+-}^{-+} ) ,
\end{equation}
where any of the terms on the left-hand side can be omitted.  We note
that the submatrix of $M$ formed by the 1st, 2nd, 4th and 5th rows and
columns has eigenvalues given in analytic form similar to
(\ref{transv-eigenvals}), as well as the submatrix formed by the 1st,
2nd, 3rd and 6th rows and columns.  This provides inequalities similar
to (\ref{transv-pos-2}) which involve different cross sections and
interference terms.

\begin{figure}
\begin{center}
\leavevmode 
\includegraphics[width=0.22\textwidth]{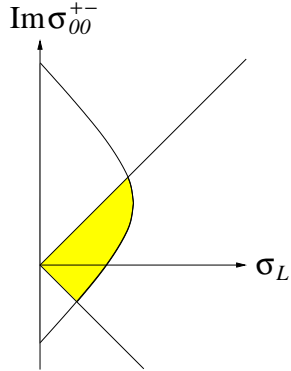}
\end{center}
\caption{\label{fig:bounds} Region in the plane of $\sigma_{L}$ and
$\im\sigma_{00}^{+-}$ allowed by the positivity bounds
(\protect\ref{long-pos}) and (\protect\ref{hyp-bound}).}
\end{figure}

As already mentioned, the dependence of the polarized $\ell p$ cross
section on $\phi$ and $\phi_S$ allows one to separate all $\gamma^* p$
cross sections and interference terms, except for
\begin{equation}
  \label{eps-combinations}
\sigma_\varepsilon^{++} 
  = \half (\sigma_{++}^{++} + \sigma_{++}^{--})
    + \varepsilon \sigma_{00}^{++} 
  = \sigma_{T} + \varepsilon \sigma_{L} , 
\qquad\qquad
\im\sigma_\varepsilon^{+-} 
     = \im( \sigma_{++}^{+-} + \varepsilon \sigma_{00}^{+-} ) ,
\end{equation}
whose individual contributions from transverse and longitudinal
photons can only be disentangled by the Rosenbluth technique.  Let us
show how the bounds (\ref{transv-pos-2}) restrict the longitudinal
contributions $\varepsilon \sigma_{L}$ and $\varepsilon\,
\im\sigma_{00}^{+-}$ to the measurable combinations
$\sigma_\varepsilon^{++}$ and $\im\sigma_\varepsilon^{+-}$.  For
simplicity we start from the bound (\ref{simpler-bound}) and omit the
term $\re\sigma_{++}^{+-}$, whose extraction requires measurement of
the angular dependence in a double spin asymmetry.  We then have
\begin{equation}
  \label{hyp-bound}
  ( A - \varepsilon \sigma_L ) \,
  ( B - \varepsilon \sigma_L )
- ( C - \varepsilon\, \im \sigma_{00}^{+-} )^2 \ge D ,
\qquad\qquad
\varepsilon \sigma_L \le \half (A+B) ,
\end{equation}
where
\begin{eqnarray}
A &=& \sigma_\varepsilon^{++}
  + \half (\sigma_{++}^{++} - \sigma_{++}^{--}) ,
\qquad\quad
B =  \sigma_\varepsilon^{++}
  - \half (\sigma_{++}^{++} - \sigma_{++}^{--}) ,
\qquad\quad
C = \im\sigma_\varepsilon^{+-} ,
\nonumber \\[0.2em]
D &=& (\re\sigma_{+-}^{++})^2 + (\im\sigma_{+-}^{++})^2 
    + ( \im\sigma_{+-}^{+-} )\, ( \im\sigma_{+-}^{-+} )
\end{eqnarray}
are measurable without Rosenbluth separation.  The corresponding
allowed region in the plane of $\sigma_{L}$ and $\im\sigma_{00}^{+-}$
is bounded on the right by a branch of the hyperbola defined by $(A -
\varepsilon \sigma_L) \, (B - \varepsilon \sigma_L) - (C -
\varepsilon\, \im \sigma_{00}^{+-})^2 = D$.  Together with
$|\im\sigma_{00}^{+-} | \le \sigma_{L}$ this leaves the shaded region
shown in Fig.~\ref{fig:bounds}.  Note that this region depends on
$\varepsilon$, both explicitly through the factors multiplying
$\sigma_{L}$ and $\im\sigma_{00}^{+-}$ in (\ref{hyp-bound}) and
implicitly through $\sigma_\varepsilon^{++}$ and
$\im\sigma_\varepsilon^{+-}$ in $A$, $B$ and $C$.  Stronger
restrictions on $\sigma_{L}$ and $\im\sigma_{00}^{+-}$ are obtained in
the same manner if one starts with the two bounds
(\ref{transv-pos-2}), each of which can be written in the form
(\ref{hyp-bound}) with suitable coefficients $A$, $B$, $C$, $D$.

\section{Deeply virtual Compton scattering}
\label{sec:dvcs}

In this section we discuss the specific case of DVCS, which is
measured in the exclusive electroproduction process
\begin{equation}
  \label{photon-proc}
\ell(l) + p(P) \to \ell(l') + \gamma(q') + p(P') ,
\end{equation}
where a real photon with momentum $q'$ now plays the role taken in the
previous sections by the produced hadron $h$ with momentum $P_h$.  We
use the same kinematical variables as before, introduced in
Sect.~\ref{sec:trafo} and in (\ref{eps-def}) and (\ref{excl-vars}).
In particular, the azimuthal angle $\phi$ is defined as in
Fig.~\ref{fig:coordinates} with $P_h$ replaced by $q'$.

\begin{figure}
\begin{center}
\leavevmode 
\includegraphics[width=0.8\textwidth]{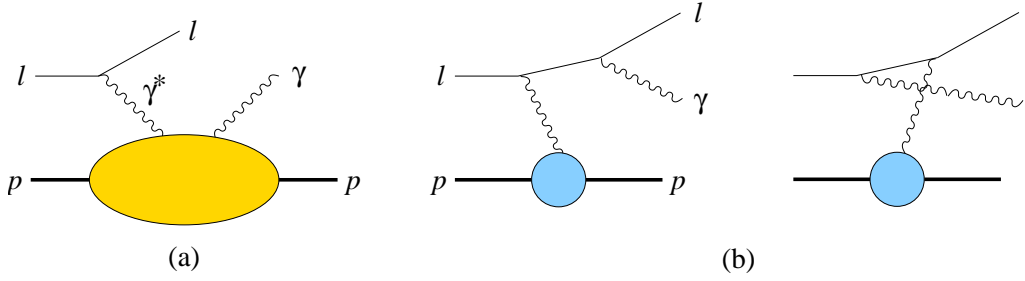}
\end{center}
\caption{\label{fig:dvcs} Graphs for virtual Compton scattering (a)
  and for the Bethe-Heitler process (b).}
\end{figure}

DVCS is one of the most valuable sources of information about
generalized parton distributions.  One reason is that in the reaction
(\ref{photon-proc}) Compton scattering interferes with the
Bethe-Heitler process, see Fig.~\ref{fig:dvcs}.  The $\ell p$ cross
section thus receives contributions
\begin{equation}
d\sigma(\ell p\to \ell \gamma p) 
  = d\sigma_{\it VCS} + d\sigma_{\it BH} + d\sigma_{\it INT}
\end{equation}
from Compton scattering and from the Bethe-Heitler process, as well as
from their interference term.  The Compton part $d\sigma_{\it VCS}$ of
the cross section has the same general structure as discussed in
Sect.~\ref{sec:ep2gp}.  With suitable kinematics and observables, one
can however also access the interference term $d\sigma_{\it INT}$,
which has a simple linear dependence on the helicity amplitudes of the
subprocess $\gamma^* p\to \gamma p$ (as opposed to a quadratic
dependence in $d\sigma_{\it VCS}$).  In addition, the interference
term provides access to the phases of these subprocess amplitudes.

In the generalized Bjorken limit of large $Q^2$ at fixed $x_B$ and
$t$, the Compton amplitude can be written as the convolution of
hard-scattering kernels with GPDs \cite{Collins:1998be}.  The detailed
dependence of the $\ell p$ cross section on these convolutions has
been given in \cite{Belitsky:2001ns}\footnote{Note that the angles
used in \protect\cite{Belitsky:2001ns} are related to the ones used
here by $\phi_{\mbox{\tiny\protect\cite{Belitsky:2001ns}}} = [\pi -
\phi \,]_{\mbox{\tiny here}}$ and
$\varphi_{\mbox{\tiny\protect\cite{Belitsky:2001ns}}} = [\pi -
\phi_{S} + \phi \,]_{\mbox{\tiny here}}\,$.}
at the leading and first subleading order in $1/Q$.  To see which
combinations of GPDs are measurable with which polarization, we give
here the expression of the interference term at leading order in
$1/Q$,
\begin{eqnarray}
  \label{dvcs-int}
\frac{d\sigma_{\it INT}}{dx_B\, dQ^2\, dt\, d\phi\, d\psi} 
&\approx &
{}- e_\ell\, \frac{\alpha_{\mathrm{em}}^3}{2\pi^2}\, 
  \frac{y^2}{Q^4}\,
  \frac{2-x_B}{|t|}\, \frac{M_p}{Q}\,
  \sqrt{\frac{2}{\varepsilon (1-\varepsilon)}}\;
  \frac{1}{P(\cos\phi)}
\nonumber \\
&& \times \Big( \cos\phi\, \re \widehat{M}_{++}
 + P_\ell \sqrt{1-\varepsilon^2}\, \sin\phi\, \im \widehat{M}_{++}
   \phantom{\Big[ \Big]}
\nonumber \\[0.2em]
&& {}+ S_L \Big[ 
 \sin\phi\, \im \widehat{M}^{\,L}_{++}
 + P_\ell \sqrt{1-\varepsilon^2}\, \cos\phi\, \re \widehat{M}^{\,L}_{++}
\Big]
\nonumber \\[0.2em]
&& {}+ S_T \cos(\phi-\phi_S)\, \Big[ 
 \sin\phi\, \im \widehat{M}^{\,S}_{++}
 + P_\ell \sqrt{1-\varepsilon^2}\, \cos\phi\, \re \widehat{M}^{\,S}_{++}
\Big] 
\nonumber \\[0.2em]
&& {}+ S_T \sin(\phi-\phi_S)\; \Big[
 \cos\phi\, \im \widehat{M}^{\,N}_{++}
 - P_\ell \sqrt{1-\varepsilon^2}\, \sin\phi\, \re \widehat{M}^{\,N}_{++}
\Big] \, \Big) \, ,
\hspace{2em}
\end{eqnarray}
where $e_\ell =\pm 1$ is the charge of the lepton beam.  Notice that
the factor
\begin{equation}
P(\cos\phi) = 1 - 2 \cos\phi \,
        \sqrt{\frac{2 (1+\varepsilon)}{\varepsilon}\,
              \frac{1-\xi}{1+\xi}\, \frac{t_0-t}{Q^2}} \,
      + O\Big(\frac{1}{Q^2}\Big)
\end{equation}
from the lepton propagators in the Bethe-Heitler amplitude influences
the $\phi$ dependence of the cross section.  This effect is formally
of order $1/Q$ but can be rather important in experimentally relevant
kinematics, also because $P(\cos\phi)$ appears in the denominator of
(\ref{dvcs-int}).  The coefficients appearing in (\ref{dvcs-int}) are
linear combinations of $\gamma^* p\to \gamma p$ helicity amplitudes
with both photons having helicity $+1$.  They can be written in terms
of Compton form factors as
\begin{eqnarray}
  \label{int-gpds}
\widehat{M}_{++} &=& \sqrt{1-\xi^2}\, \frac{\sqrt{t_0-t}}{2M_p}\, 
 \Bigg[ F_1 \mathcal{H} + \xi (F_1 + F_2) \tilde{\mathcal{H}}
                        - \frac{t}{4M_p^2}\, F_2\, \mathcal{E} 
 \Bigg] ,
\\
\widehat{M}_{++}^{\,L} &=& 
 \sqrt{1-\xi^2}\, \frac{\sqrt{t_0-t}}{2M_p}\,
 \Bigg[ F_1 \tilde\mathcal{H}
      + \xi (F_1 + F_2) \Big( {\mathcal{H}} 
      + \frac{\xi}{1+\xi}\, \mathcal{E} \Big)
      - \Big( \frac{\xi}{1+\xi}\, F_1 + \frac{t}{4M_p^2}\, F_2 \Big)\,
             \xi \tilde\mathcal{E} \Bigg] ,
\nonumber \\
\widehat{M}^{\,S}_{++} &=&
  \Bigg[ \xi^2 \Big( F_1 + \frac{t}{4M_p^2}\, F_2 \Big)
         - \frac{t}{4M_p^2}\, F_2 \Bigg] \tilde\mathcal{H}
    - \Big( \frac{t}{4M_p^2} + \frac{\xi^2}{1+\xi} \Big)\,
	\xi (F_1+F_2) \, \mathcal{E}
\nonumber \\
 && {}+ \Bigg[ 
 \Big( \frac{t}{4M_p^2} + \frac{\xi^2}{1+\xi} \Big)\, F_1
     + \frac{t}{4M_p^2} \, \xi F_2 \Bigg]\, \xi \tilde\mathcal{E}
     - \xi^2 (F_1+F_2) \mathcal{H} \, ,
\nonumber \\[0.2em]
\widehat{M}^{\,N}_{++} &=&
{}- \frac{t}{4M_p^2}\, \Big( F_2 \mathcal{H} - F_1\, \mathcal{E} \Big)
  + \xi^2 \Big( F_1 + \frac{t}{4M_p^2}\, F_2 \Big) \,
	( \mathcal{H} + \mathcal{E} )
  - \xi^2 (F_1+F_2) \Big( \tilde{\mathcal{H}} 
        + \frac{t}{4M_p^2}\, \tilde{\mathcal{E}} \Big) ,
\nonumber
\end{eqnarray}
with the Dirac and Pauli form factors $F_1$ and $F_2$ of the proton
evaluated at momentum transfer $t$.  The term with superscript $S$
(``sideways'') contributes most strongly to the cross section for
transverse target polarization in the hadron plane, and the term with
superscript $N$ (``normal'') contributes most for target polarization
perpendicular to the hadron plane, according to the respective factors
$\cos(\phi-\phi_S)$ and $\sin(\phi-\phi_S)$ in (\ref{dvcs-int}).  The
Compton form factors are given as integrals over GPDs and read
\begin{eqnarray}
  \label{compton-integrals}
\mathcal{H}(\xi,t) &=& \sum_q e_q^2
\int_{-1}^1 dx\, H^q(x,\xi,t)\, 
    \left( \frac{1}{\xi - x - i\varepsilon}
           - \frac{1}{\xi + x - i\varepsilon} \right) + O(\alpha_s) ,
\nonumber \\
\tilde{\mathcal{H}}(\xi,t) &=& \sum_q e_q^2
\int_{-1}^1 dx\, \tilde{H}^q(x,\xi,t)\, 
    \left( \frac{1}{\xi - x - i\varepsilon}
           + \frac{1}{\xi + x - i\varepsilon} \right) + O(\alpha_s) ,
\end{eqnarray}
where the sums are over quark flavors $q$ with $e_u= \frac{2}{3}$ and
$e_d= e_s= -\frac{1}{3}$.  The expressions for $\mathcal{E}$ and
$\smash{\tilde{\mathcal{E}}}$ are analogous to those of $\mathcal{H}$
and $\smash{\tilde{\mathcal{H}}}$, respectively.  In
(\ref{compton-integrals}) we have suppressed the dependence of the
Compton form factors on $Q^2$, which arises at order $\alpha_s$ in
analogy to the scaling violation in deep inelastic structure
functions.

We see in (\ref{dvcs-int}) that single beam or target spin asymmetries
project out the imaginary parts of the Compton form factors, which
according to (\ref{compton-integrals}) are just GPDs at $x=\pm \xi$ to
leading order in $\alpha_s$.  The real part of $\widehat{M}_{++}$
appears in the unpolarized $\ell p$ cross section, and the real parts
of the other three combinations in double spin asymmetries.  {}From
(\ref{int-gpds}) one readily finds that separation of all four Compton
form factors is possible.

Since $\xi$ is small in a wide range of experimentally relevant
kinematics, it is instructive to write
\begin{eqnarray}
  \label{int-gpds-sim}
\widehat{M}^{\,S}_{++} &=&
- \frac{t}{4M_p^2}\, \Big[ F_2 \tilde\mathcal{H} 
                         - F_1\, \xi \tilde\mathcal{E}\, \Big]
  - \xi \frac{t}{4M_p^2}\, \Big[ (F_1+F_2) \mathcal{E} 
                               - F_2\, \xi \tilde\mathcal{E}\, \Big]
  + \xi^2\, O( \mathcal{H}, \mathcal{E}, \tilde\mathcal{H},
                 \xi \tilde\mathcal{E} ) \, ,
\nonumber \\
\widehat{M}^{\,N}_{++} &=&
- \frac{t}{4M_p^2}\, \Big[ F_2 \mathcal{H} - F_1 \mathcal{E} \Big]
  - \xi \frac{t}{4M_p^2}\, (F_1+F_2)\, \xi \tilde\mathcal{E}
  + \xi^2\, O( \mathcal{H}, \mathcal{E}, \tilde\mathcal{H} ) \, .
\end{eqnarray}
For counting powers of $\xi$ we use $\xi\tilde\mathcal{E}$ rather than
$\tilde\mathcal{E}$ in comparison with $\mathcal{H}$, because the
contribution to $\tilde{E}$ from pion exchange scales like $\xi^{-1}$,
see \cite{Goeke:2001tz,Diehl:2003ny}.  The only combination in
(\ref{int-gpds}) where the helicity-flip distribution $E$ is not
kinematically suppressed compared with other GPDs turns out to be
$\widehat{M}^{\,N}_{++}$, which comes with an angular dependence like
$\sin(\phi-\phi_S) \cos\phi$ or $\sin(\phi-\phi_S) \sin\phi$ in the
interference term.  Note that one may rewrite
\begin{eqnarray}
\lefteqn{
\phantom{+} \cos(\phi-\phi_S)\, \Big[ 
 \sin\phi\, \im \widehat{M}^{\,S}_{++}
 + P_\ell \sqrt{1-\varepsilon^2}\, \cos\phi\, \re \widehat{M}^{\,S}_{++}
\Big] 
}
\nonumber \\
\lefteqn{
+ \sin(\phi-\phi_S)\; \Big[
 \cos\phi\, \im \widehat{M}^{\,N}_{++}
 - P_\ell \sqrt{1-\varepsilon^2}\, \sin\phi\, \re \widehat{M}^{\,N}_{++}
\Big]
}
\nonumber \\
&=& \frac{1}{2} \Big[
  \sin(2\phi - \phi_S)\, 
      \im (\widehat{M}^{\,S}_{++} + \widehat{M}^{\,N}_{++})
+ P_\ell \sqrt{1-\varepsilon^2}\, \cos(2\phi - \phi_S)\, 
      \re (\widehat{M}^{\,S}_{++} + \widehat{M}^{\,N}_{++})
\nonumber \\
&& \hspace{3.07em} {}+ \sin\phi_S\, 
      \im (\widehat{M}^{\,S}_{++} - \widehat{M}^{\,N}_{++})
+ P_\ell \sqrt{1-\varepsilon^2}\, \cos\phi_S\, 
      \re (\widehat{M}^{\,S}_{++} - \widehat{M}^{\,N}_{++}) \,\Big] \, ,
\end{eqnarray}
which results in a simpler form of the angular dependence, as we have
used in Sect.~\ref{sec:ep2gp}.  In terms of dominant contributions
from the different GPDs, the combinations $\widehat{M}^{\,S}_{++}$ and
$\widehat{M}^{\,N}_{++}$ appear however more natural than their
difference and sum, see (\ref{int-gpds-sim}).

Let us now take a closer look at how the different Compton form
factors can be extracted from the polarized $\ell p$ cross section.
To this end we need the general dependence on the angles $\phi$ and
$\phi_S$, which has the form \cite{Belitsky:2001ns}
\begin{eqnarray}
  \label{compton-gen}
\frac{Q^4}{y^2}\,
\frac{d\sigma_{\it BH}}{dx_B\, dQ^2\, dt\, d\phi\, d\psi} &=& 
  \frac{1}{|t|}\, \frac{1}{\varepsilon}\, \frac{1}{P(\cos\phi)} \Bigg(
    \sum_{n=0}^2 \cos(n\phi)\, c_{nU}^{\it BH}
  + S_L P_\ell\, \sum_{n=0}^1 \cos(n\phi)\, c_{nL}^{\it BH}
\\
 && {}+ S_T P_\ell\, \Big[ \cos(\phi-\phi_S) 
                     \sum_{n=0}^1 \cos(n\phi)\, c_{nS}^{\it BH} 
      + \sin(\phi-\phi_S) \sin\phi\; s_{1N}^{\it BH} \,\Big] \Bigg) ,
\nonumber \\
\frac{Q^4}{y^2}\,
\frac{d\sigma_{\it VCS}}{dx_B\, dQ^2\, dt\, d\phi\, d\psi} &=&
  \frac{1}{Q^2}\, \frac{1}{1-\varepsilon} \Bigg(
    \sum_{n=0}^2 \cos(n\phi)\, c_{nU}^{\it VCS}
  + P_\ell \sin\phi\; s_{1U}^{\it VCS}
\nonumber \\
 && {}+ S_L\, \sum_{n=1}^2 \sin(n\phi)\, s_{nL}^{\it VCS}
      + S_L P_\ell\, \sum_{n=0}^1 \cos(n\phi)\, c_{nL}^{\it VCS}
\nonumber \\
 && {}+ S_T\, \Big[
    \sin(\phi-\phi_S) \sum_{n=0}^2 \cos(n\phi)\, c_{nN}^{\it VCS}
  + \cos(\phi-\phi_S) \sum_{n=1}^2 \sin(n\phi)\, s_{nS}^{\it VCS} 
\,\Big]
\nonumber \\
 && {}+ S_T P_\ell\, \Big[
    \cos(\phi-\phi_S) \sum_{n=0}^1 \cos(n\phi)\, c_{nS}^{\it VCS}
  + \sin(\phi-\phi_S) \sin\phi\; s_{1N}^{\it VCS} \,\Big] \Bigg) ,
\nonumber \\
\frac{Q^4}{y^2}\,
\frac{d\sigma_{\it INT}}{dx_B\, dQ^2\, dt\, d\phi\, d\psi} &=&
  {}- e_\ell\, \frac{1}{|t|}\,
  \frac{M_p}{Q}\, \frac{1}{\varepsilon \sqrt{1-\varepsilon}}\,
  \frac{1}{P(\cos\phi)} \Bigg(
  \sum_{n=0}^3 \cos(n\phi)\, c_{nU}^{\it INT} +
  P_\ell \sum_{n=1}^2 \sin(n\phi)\, s_{nU}^{\it INT}
\nonumber \\
 && {}+ S_L \sum_{n=1}^3 \sin(n\phi)\, s_{nL}^{\it INT}
  + S_L P_\ell \sum_{n=0}^2 \cos(n\phi)\, c_{nL}^{\it INT}
\nonumber \\
 && {}+ S_T \Big[ 
    \sin(\phi-\phi_S) \sum_{n=0}^3 \cos(n\phi)\, c_{nN}^{\it INT}
  + \cos(\phi-\phi_S) \sum_{n=1}^3 \sin(n\phi)\, s_{nS}^{\it INT} 
\,\Big]
\nonumber \\
 && {}+ S_T P_\ell \Big[ 
    \cos(\phi-\phi_S) \sum_{n=0}^2 \cos(n\phi)\, c_{nS}^{\it INT}
  + \sin(\phi-\phi_S) \sum_{n=1}^2 \sin(n\phi)\, s_{nN}^{\it INT} 
\,\Big] \Bigg) ,
\nonumber 
\end{eqnarray}
where the subscripts $U$, $L$, $S$, $N$ of the angular coefficients
$c$ and $s$ indicate an unpolarized target, or longitudinal, sideways
or normal target polarization as explained after (\ref{int-gpds}).
These angular coefficients depend on $\varepsilon$, $x_B$, $Q^2$, $t$,
and the kinematic prefactors have been chosen such that (up to
logarithms in $Q^2$) they all remain finite or vanish in the limit of
large $Q^2$ relevant for the extraction of GPDs.\footnote{For the
purpose of our presentation we have normalized the coefficients $c$,
$s$ differently than in \protect\cite{Belitsky:2001ns}, and we have
chosen a different notation to indicate the target spin dependence.}
For $\varepsilon\to 0$ the coefficients behave like $c_n, s_n \sim
\sqrt{\varepsilon}^{\,n}$, and for $\varepsilon\to 1$ the coefficients
accompanied by the lepton polarization $P_\ell$ vanish like
$\sqrt{1-\varepsilon}$ whereas the others remain finite.  We see in
(\ref{compton-gen}) that generically $\sigma_{\it BH}$ dominates over
$\sigma_{\it VCS}$ in generalized Bjorken kinematics (where $|t| \ll
Q^2$) except if $\varepsilon$ is sufficiently close to 1.  The
interference term lies in between $\sigma_{\it BH}$ and $\sigma_{\it
VCS}$, and it can most directly be isolated from the difference of
cross sections for positive and negative lepton beam charge.
Furthermore, we see that the Bethe-Heitler contribution depends only
on the \emph{product} of beam and target polarizations, so that it
drops out in single beam or target spin asymmetries.  Unless
$\varepsilon$ is close to 1 these asymmetries will then be dominated
by the interference term, with smaller contributions from the Compton
cross section.

Let us now discuss the dynamical content and the power behavior in $Q$
of the angular coefficients in the generalized Bjorken limit.  It is
independent of the target polarization, and in the following we write
$c_n$ and $s_n$ to collectively denote the coefficients with
subscripts $U$, $L$, $S$, $N$.  Detailed formulae and references can
be found in \cite{Belitsky:2001ns}.  It is understood that the power
behavior discussed in the following is modified by logarithms in $Q^2$
for the Compton and interference terms.
\begin{enumerate}
\item The Bethe-Heitler coefficients $c_n^{\it BH}$ and $s_n^{\it BH}$
  behave like $1/Q^n$.
\item The leading coefficients in the Compton cross section are the
  $c_0^{\it VCS}$, which (up to logarithms) become independent of $Q$
  in the Bjorken limit.  They are quadratic in the twist-two Compton
  form factors $\mathcal{H}$, $\mathcal{E}$,
  $\smash{\tilde{\mathcal{H}}}$, $\smash{\tilde{\mathcal{E}}}$
  introduced above, which parameterize $\gamma^* p\to \gamma p$
  amplitudes with equal helicity of the initial and final state
  photon.

  $c_1^{\it VCS}$ and $s_1^{\it VCS}$ are suppressed by $1/Q$ and can
  be expressed through products of twist-two with twist-three Compton
  form factors.  The twist-three form factors parameterize the
  $\gamma^* p\to \gamma p$ amplitudes with a longitudinal $\gamma^*$.
  They contain a part involving the twist-two GPDs already discussed
  and another part involving matrix elements of quark-antiquark-gluon
  operators in the nucleon, in analogy with the sum $g_1+g_2$ of
  inclusive structure functions for DIS.

  $c_2^{\it VCS}$ and $s_2^{\it VCS}$ become again $Q$ independent in
  the Bjorken limit, but only start at order $\alpha_s$.  They can be
  expressed through products of the Compton form factors
  $\mathcal{H}$, $\mathcal{E}$, $\smash{\tilde{\mathcal{H}}}$,
  $\smash{\tilde{\mathcal{E}}}$ with form factors parameterizing
  $\gamma^* p\to \gamma p$ transitions from photon helicity $-1$ to
  $+1$.  These transitions have a twist-two contribution from gluon
  transversity distributions, coming of course with a factor of
  $\alpha_s$.  They also have a twist-four contribution from quark
  distributions, which comes without $\alpha_s$ but with a $1/Q^2$
  suppression \cite{Kivel:2001rw}.  Very little is known about gluon
  transversity distributions, so that we cannot say which piece will
  be more important in given kinematics.
\item In the interference term the leading coefficients are $c_1^{\it
  INT}$ and $s_1^{\it INT}$, as we already saw in (\ref{dvcs-int}).
  They provide access to the linear form factor combinations
  (\ref{int-gpds}) and thus are especially important observables to
  extract from measurement.

  The coefficients $c_0^{\it INT}$ involve the Compton form factors
  $\mathcal{H}$, $\mathcal{E}$, $\smash{\tilde{\mathcal{H}}}$,
  $\smash{\tilde{\mathcal{E}}}$ as well, but they come with a
  kinematical suppression factor $1/Q$.

  The coefficients $c_2^{\it INT}$ and $s_2^{\it INT}$ are linear
  combinations of twist-three Compton form factors and scale as $1/Q$.
  If one is willing to make the Wandzura-Wilczek approximation, where
  quark-antiquark-gluon matrix elements are neglected, these
  observables provide additional information on the twist-two
  distributions $H$, $E$, $\tilde{H}$, $\tilde{E}$.
  
  $c_3^{\it INT}$ and $s_3^{\it INT}$ are sensitive to the $\gamma^*
  p\to \gamma p$ transitions from photon helicity $-1$ to $+1$ and
  thus have a $Q$ independent piece starting at order $\alpha_s$.
\end{enumerate}
For completeness we remark that the angular coefficients $c^{\it INT}$
and $s^{\it INT}$ of the interference term receive further
contributions \cite{Diehl:2003ny}, which are suppressed compared with
those just discussed by either powers of $1/Q^2$ or of $\alpha_s$.  We
need not discuss them here, given the accuracy we aim at.

Using the transformation rules from Sects.~\ref{sec:long} and
\ref{sec:transv} and some relations between trigonometric functions,
one can readily extract from (\ref{compton-gen}) the $\ell p$ cross
sections for longitudinal and for transverse target polarization with
respect to the beam, as we did in Sect.~\ref{sec:ep2gp}.  We restrict
ourselves here to an unpolarized lepton beam, where the Bethe-Heitler
cross section does not contribute to the $P_L$ or $P_T$ dependence as
mentioned above.  In suitable kinematics one is then most sensitive to
the interference term, which reads
\begin{eqnarray}
  \label{int-long}
\lefteqn{
P(\cos\phi)\;
\frac{d\sigma_{\it INT}}{dx_B\, dQ^2\, dt\,
    d\phi} \; \Bigg|_{P_T=0, P_\ell=0} 
 \propto\; \mbox{terms independent of $P_L$} 
}
\nonumber \\
 &+& P_L \Bigg(
   \sin\phi\, \Big[ \cos\theta\, s_{1L}^{\it INT} 
     - \sin\theta\, c_{0N}^{\it INT}
     + \half\sin\theta\, (c_{2N}^{\it INT} - s_{2S}^{\it INT}) \Big]
\nonumber \\
 && \hspace{0.7em} 
   {}+ \sin(2\phi)\, \Big[ \cos\theta\, s_{2L}^{\it INT}
     - \half\sin\theta\, (c_{1N}^{\it INT} + s_{1S}^{\it INT})
     + \half\sin\theta\, (c_{3N}^{\it INT} - s_{3S}^{\it INT}) \Big]
\phantom{\Bigg( \Bigg) }
\nonumber \\
 && \hspace{0.7em} 
   {}+  \sin(3\phi)\, \Big[ \cos\theta\, s_{3L}^{\it INT}
     - \half\sin\theta\, (c_{2N}^{\it INT} + s_{2S}^{\it INT}) \Big]
   - \sin(4\phi)\, \half\sin\theta\, (c_{3N}^{\it INT} + s_{3S}^{\it INT}) 
\,\Bigg)
\hspace{3em}
\end{eqnarray}
for longitudinal and
\begin{eqnarray}
  \label{int-transv}
\lefteqn{
\frac{(1 - \sin^2\!\theta\, 
              \sin^2\!\phi_S)^{3/2} \rule{0pt}{1.1em}}{\cos\theta}\,
P(\cos\phi)\;
\frac{d\sigma_{\it INT}}{dx_B\, dQ^2\, dt\, d\phi\,
    d\phi_S} \; \Bigg|_{P_L=0, P_\ell=0} 
 \propto\; \mbox{terms independent of $P_T$} 
}
\nonumber \\
 &+& P_T \sin(\phi-\phi_S)\, \Bigg(
    \cos\theta\, c_{0N}^{\it INT} + \half\sin\theta\, s_{1L}^{\it INT}
\nonumber \\
 && \hspace{4.65em} \phantom{\Bigg( \Bigg) }
    {}+ \cos\phi\, \Big[ \cos\theta\, c_{1N}^{\it INT}
    + \half\sin\theta\, s_{2L}^{\it INT} \Big]
    + \cos(2\phi)\, \Big[ \cos\theta\, c_{2N}^{\it INT}
    - \half\sin\theta\, (s_{1L}^{\it INT} - s_{3L}^{\it INT}) \Big]
\nonumber \\
 && \hspace{6.1em} 
    {}+ \cos(3\phi)\, \Big[ \cos\theta\, c_{3N}^{\it INT}
    - \half\sin\theta\, s_{2L}^{\it INT} \Big]
    - \cos(4\phi)\, \half\sin\theta\, s_{3L}^{\it INT} \Bigg)
\nonumber \\
 &+& P_T \cos(\phi-\phi_S)\, \Bigg(
      \sin\phi\, \Big[ \cos\theta\, s_{1S}^{\it INT}
    + \half\sin\theta\, s_{2L}^{\it INT} \Big]
    + \sin(2\phi)\, \Big[ \cos\theta\, s_{2S}^{\it INT}
    + \half\sin\theta\, (s_{1L}^{\it INT} + s_{3L}^{\it INT}) \Big]
\nonumber \\
 && \hspace{6.1em} 
    {}+ \sin(3\phi)\, \Big[ \cos\theta\, s_{3S}^{\it INT}
    + \half\sin\theta\, s_{2L}^{\it INT} \Big]
    + \sin(4\phi)\, \half\sin\theta\, s_{3L}^{\it INT} \,\Bigg)
\end{eqnarray}
for transverse target polarization, where we have not displayed
kinematic factors which are independent on $\phi$ and $\phi_S$.  For
both polarizations, the $\sin\phi$ or $\cos\phi$ modulation in the
cross section (at given $\phi-\phi_S$) receives its main contribution
from the coefficients $s_{1L}^{\it INT}$, $c_{1N}^{\it INT}$ or
$s_{1S}^{\it INT}$ containing the twist-two Compton form factors, with
corrections that are power suppressed by $1/Q^2$.  In the
$\sin(2\phi)$ and $\cos(2\phi)$ terms, however, the coefficients
$s_{2L}^{\it INT}$, $c_{2N}^{\it INT}$ and $s_{2S}^{\it INT}$
containing the twist-three Compton form factors appear together with
other terms of the same order in $1/Q$.  Their extraction would
require at least subtraction of the contributions from the
coefficients $s_{1L}^{\it INT}$, $c_{1N}^{\it INT}$, $s_{1S}^{\it
INT}$, which are presumably larger than $s_{3L}^{\it INT}$,
$c_{3N}^{\it INT}$, $s_{3S}^{\it INT}$ according to our discussion
above.

A rigorous separation of $s_{1L}^{\it INT}$, $c_{1N}^{\it INT}$ and
$s_{1S}^{\it INT}$ from the $1/Q^2$ corrections that accompany them in
the $\sin\phi$ or $\cos\phi$ terms requires measurement of almost the
full $\phi$ and $\phi_S$ dependence in the polarized cross sections
(the information from the $\sin(4\phi)$ and $\cos(4\phi)$ terms is
redundant).  For small enough $\sin\theta$ one can however easily
estimate whether these $1/Q^2$ corrections are numerically important,
provided one knows the size of the $\sin(2\phi)$ term in
(\ref{int-long}) and of the $\sin(\phi-\phi_S)$, $\sin(\phi-\phi_S)
\cos(2\phi)$ and $\cos(\phi-\phi_S) \sin(2\phi)$ terms in
(\ref{int-transv}).


\section{Summary}
\label{sec:sum}

We have studied the analysis of lepton scattering on a polarized spin
$\half$ target.  Starting point was the general transformation between
target spin states defined with respect to the lepton beam direction,
which are relevant in experiment, and spin states defined with respect
to the lepton momentum transfer $\tvec{q} = \tvec{l} - \tvec{l}'$,
which are natural to describe the hadronic part of the process in the
one-photon exchange approximation.  This transformation can easily be
incorporated at the level of polarized cross sections and of spin
asymmetries.

Detailed information on spin properties of the nucleon can be obtained
in semi-inclusive and in exclusive $\ell p$ scattering from the
distribution in the azimuthal angle $\phi$ between the lepton
scattering plane and a suitably defined hadron plane.  We have given
the general form of the $\ell p$ cross section for longitudinal or
transverse target polarization relative to the beam direction,
expressed in terms of polarized cross sections and interference terms
of the $\gamma^* p$ subprocess.  Our main results, given in
(\ref{Xsection-master}), (\ref{Xsection-long}) and
(\ref{Xsection-transv}) are valid for all kinematics and thus hold in
a variety of dynamical contexts.  They can be used for any definition
of a hadronic plane, provided this definition depends only on
four-momenta of the $\gamma^* p$ subprocess.  They readily generalize
to cross sections which depend on kinematical variables describing the
hadronic final state, provided these variables are invariant under a
parity transformation.  Combining the information from both
longitudinal and transverse target polarization, one can separate all
$\gamma^* p$ cross sections and interference terms, except for the
contributions from longitudinal and transverse photons to $\sigma_T +
\varepsilon\sigma_L$ and to its counterpart $\im( \sigma_{++}^{+-} +
\varepsilon\sigma_{00}^{+-} )$ for proton helicity-flip.  These
contributions can be disentangled only by Rosenbluth separation, which
requires measurement at different $\ell p$ energies.  Without this
possibility, one can however use positivity constraints to obtain
limits on $\sigma_L$ and $\im\sigma_{00}^{+-}$ from measuring the
angular dependence of the polarized $\ell p$ cross sections.

We have then studied the particular cases of semi-inclusive deep
inelastic scattering and of exclusive meson production.  We have also
considered the case of deeply virtual Compton scattering, where a
special angular and polarization dependence arises from the
interference term between Compton scattering and the Bethe-Heitler
process.  Taking into account the power behavior in the large scale
$Q$ for each of these reactions, we have in particular discussed how
from measured cross sections one can separate twist-two and
twist-three quantities, whose analysis in QCD provides specific
information on the role of spin at the interface of partons and
hadrons.

The parameter controlling the mixing of polarizations defined relative
to the beam or to the photon direction is $\gamma = 2 x_B M_p /Q$.
For deep inelastic measurements at low $x_B$ one can thus typically
neglect this mixing and directly use cross section formulae like
(\ref{Xsection-master}) and (\ref{compton-gen}) for the analysis.  For
moderate or high $x_B$, our results allow one to take these mixing
effects into account without further model assumptions.


\section*{Acknowledgments} 

We gratefully acknowledge discussions with D. Boer, J. Collins,
P. Mulders, and with many of our colleagues from the HERMES
collaboration.  Special thanks go to D.~Hasch, O.~Nachtmann and
W.-D.~Nowak for valuable remarks on the manuscript.  The work of M.D.\
is supported by the Helmholtz Assciation, contract number VH-NG-004.
S.S.\ acknowledges support by the DESY Summer Student Programme and
thanks DESY for warm hospitality.


\appendix
\section{Interference terms vs.\ cross sections}
\label{app:inter-cross}

Interference terms $\sigma_{mn}^{ij}$ between different polarizations
in the process $\gamma^* p \to h X$ can be expressed through cross
sections in a suitable basis of spin states.  In particular, we have
\begin{eqnarray}
\re \sigma_{++}^{+-} &=& \half \Big( \,
    \sigma_{++}^{\rightarrow\rightarrow} 
  - \sigma_{++}^{\leftarrow\leftarrow} \, \Big) ,
\nonumber \\
\im \sigma_{++}^{+-} &=& -\half \Big( \,
    \sigma_{++}^{\uparrow\uparrow} 
  - \sigma_{++}^{\downarrow\downarrow} \, \Big) ,
\qquad\qquad
\im \sigma_{00}^{+-} \;=\; -\half \Big( \,
    \sigma_{00}^{\uparrow\uparrow} 
  - \sigma_{00}^{\downarrow\downarrow} \; \Big) ,
\end{eqnarray}
where the labels $\rightarrow$ and $\leftarrow$ respectively denote
definite proton spin projection $+\frac{1}{2}$ and $-\frac{1}{2}$
along the $x''$ axis, and the labels $\uparrow$ and $\downarrow$
definite proton spin projection $+\frac{1}{2}$ and $-\frac{1}{2}$
along the $y''$ axis.  In other words, $\re \sigma_{++}^{+-}$
corresponds to the asymmetry for transverse proton polarization
\emph{in} the hadron plane, and $\im \sigma_{++}^{+-}$ and $\im
\sigma_{00}^{+-}$ to asymmetries for transverse proton polarization
\emph{normal to} the hadron plane.

The interference terms between photon helicities $+1$ and $-1$ can be
written as combinations of cross sections for linear photon
polarization.  With photon polarization vectors
\begin{equation}
\epsilon_\rightarrow = (0,1,0,0) , \qquad
\epsilon_\uparrow = (0,0,1,0) , \qquad
\epsilon_\nearrow = \frac{1}{\sqrt{2}}\, (0,1,1,0) , \qquad
\epsilon_\nwarrow = \frac{1}{\sqrt{2}}\, (0,-1,1,0)
\end{equation}
defined in coordinate system $C''$ we have
\begin{eqnarray}
\re \sigma_{+-}^{++} &=& \half \Big( \,
    \sigma_{\uparrow\uparrow}^{++} 
  - \sigma_{\rightarrow\rightarrow}^{++} \, \Big) ,
\nonumber \\
\im \sigma_{+-}^{++} &=& \half \Big( \,
    \sigma_{\nearrow\!\!\nearrow}^{++} 
  - \sigma_{\,\nwarrow\!\!\nwarrow}^{++} \, \Big) ,
\nonumber \\
\im (\sigma_{+-}^{+-} + \sigma_{+-}^{-+}) &=& \half \Big( \,
    \sigma_{\nearrow\!\!\nearrow}^{\rightarrow\rightarrow} 
  - \sigma_{\nearrow\!\!\nearrow}^{\leftarrow\leftarrow} 
  - \sigma_{\,\nwarrow\!\!\nwarrow}^{\rightarrow\rightarrow} 
  + \sigma_{\,\nwarrow\!\!\nwarrow}^{\leftarrow\leftarrow} \, \Big) ,
\nonumber \\
\im (\sigma_{+-}^{+-} - \sigma_{+-}^{-+}) &=& {}- \half \Big( \,
    \sigma_{\uparrow\uparrow}^{\uparrow\uparrow} 
  - \sigma_{\uparrow\uparrow}^{\downarrow\downarrow} 
  - \sigma_{\rightarrow\rightarrow}^{\uparrow\uparrow} 
  + \sigma_{\rightarrow\rightarrow}^{\downarrow\downarrow} \, \Big) .
\end{eqnarray}


\section{Inclusive deep inelastic scattering}
\label{app:usual-dis}

Our derivation in Sect.~\ref{sec:ep2gp} can readily be adapted to
inclusive lepton-proton scattering $\ell p\to \ell X$.  The inclusive
hadronic state $X$ does not define a hadron plane, so that we
introduce $\gamma^* p$ cross sections and interference terms for
photon and proton polarizations with respect to the lepton plane
spanned by $\tvec{q}$ and $\tvec{l}'$ in the target rest frame (cf.\
also our remarks at the end of Sect.~\ref{sec:dihadron}).  In the
inclusive case we have additional symmetry relations $\sigma_{nm}^{ji}
= \sigma^{ij}_{mn}$ since the inclusive hadronic tensor is constrained
by time reversal invariance.\footnote{For the semi-inclusive or
exclusive case time reversal does not constrain the hadronic tensor
(\ref{had-tensor}) since it transforms the states $|h X\rangle$ from
``out'' to ``in'' states.  In the inclusive case this is of no
consequence because one sums over a \emph{complete} set of final
states.}
We then obtain for the $\ell p$ cross section
\begin{eqnarray}
  \label{incl-long}
\lefteqn{
\Bigg[ \frac{\alpha_{\rm em}}{2\pi}\, \frac{y^2}{1-\varepsilon}\,
       \frac{1-x_B}{x_B}\, \frac{1}{Q^2} \Bigg]^{-1}
\frac{d\sigma}{dx_B\, dQ^2} \; \Bigg|_{P_T=0} 
  = \frac{1}{2} \Big( \sigma_{++}^{++} + \sigma_{++}^{--} \Big)
    + \varepsilon \sigma_{00}^{++}
}
\nonumber \\
 &+& P_L P_\ell\, \Bigg[
   \cos\theta \sqrt{1-\varepsilon^2}\; \frac{1}{2} 
   ( \sigma_{++}^{++} - \sigma_{++}^{--} )
 + \sin\theta \sqrt{\varepsilon (1-\varepsilon)}\, 
   \sigma_{+0}^{+-} \Bigg]
\end{eqnarray}
for longitudinal and
\begin{eqnarray}
  \label{incl-transv}
\lefteqn{
\Bigg[ \frac{\alpha_{\rm em}}{4\pi^2}\, \frac{y^2}{1-\varepsilon}\,
       \frac{1-x_B}{x_B}\, \frac{1}{Q^2} \Bigg]^{-1}
\frac{d\sigma}{dx_B\, dQ^2\, d\psi} \; \Bigg|_{P_L=0} 
  =  \frac{1}{2} \Big( \sigma_{++}^{++} + \sigma_{++}^{--} \Big)
     + \varepsilon \sigma_{00}^{++} 
}
\nonumber \\
 &-& P_T P_\ell\, \cos\psi\, \Bigg[
   \cos\theta \sqrt{\varepsilon (1-\varepsilon)}\, \sigma_{+0}^{+-}
 - \sin\theta \sqrt{1-\varepsilon^2}\; \frac{1}{2} 
   ( \sigma_{++}^{++} - \sigma_{++}^{--} ) \, \Bigg]
\end{eqnarray}
for transverse target polarization.  Using the relation between $\psi$
and $\phi_S$ from Sect.~\ref{sec:transv} and taking into account that
$\sigma_{+0}^{+-}$ is now purely real because of time reversal
invariance, we see that this corresponds to the $\phi$ independent
terms of our formulae (\ref{Xsection-master}), (\ref{Xsection-long})
and (\ref{Xsection-transv}) for semi-inclusive or exclusive
scattering.  It is customary to introduce double spin asymmetries
\cite{Adams:1994id}
\begin{eqnarray}
A_\parallel(x_B, y, Q^2) &=& 
  \frac{d\sigma^\rightarrow(P_L=+1) - d\sigma^\rightarrow(P_L=-1)
      - d\sigma^\leftarrow(P_L=+1)  + d\sigma^\leftarrow(P_L=-1)}{
        d\sigma^\rightarrow(P_L=+1) + d\sigma^\rightarrow(P_L=-1)
      + d\sigma^\leftarrow(P_L=+1)  + d\sigma^\leftarrow(P_L=-1)}
  \; \Bigg|_{P_T=0} \, ,
\nonumber \\[0.3em]
A_\perp(x_B, y, Q^2) &=& {}- \frac{1}{\cos\psi}\;
  \frac{d\sigma^\rightarrow(\psi) - d\sigma^\rightarrow(\psi+\pi)
      - d\sigma^\leftarrow(\psi)  + d\sigma^\leftarrow(\psi+\pi)}{
        d\sigma^\rightarrow(\psi) + d\sigma^\rightarrow(\psi+\pi)
      + d\sigma^\leftarrow(\psi)  + d\sigma^\leftarrow(\psi+\pi)}
  \; \Bigg|_{P_T=1, P_L=0}
\end{eqnarray}
where $\sigma^\rightarrow$ denotes right-handed and
$\sigma^\leftarrow$ left-handed lepton beam polarization.  Note that
the $\psi$ dependence of the numerator is divided out in $A_\perp$.
One further introduces asymmetries $A_1$ and $A_2$ for the subprocess
$\gamma^* p\to X$, which are related to the usual inclusive structure
functions by
\begin{equation}
A_1(x_B, Q^2) = \frac{g_1 - \gamma^2 g_2}{F_1} , \qquad\qquad
A_2(x_B, Q^2) = \frac{\gamma (g_1 + g_2)}{F_1} .
\end{equation}
The relation between lepton and photon asymmetries is usually
given in the form \cite{Filippone:2001ux}
\begin{equation}
A_\parallel = D\, (A_1 + \eta A_2) , \qquad\qquad
A_\perp     = d\, (A_2 - \zeta A_1) ,
\end{equation}
where
\begin{equation}
D = \frac{1-(1-y) \varepsilon}{1 + \varepsilon R} , \qquad
\eta = \frac{\gamma y \varepsilon}{1-(1-y) \varepsilon} , \qquad
d = D \sqrt{\frac{2\varepsilon}{1+\varepsilon}} , \qquad
\zeta = \eta\, \frac{1+\varepsilon}{2\varepsilon}
\end{equation}
with $R = \sigma_L /\sigma_T$.  In our notation this reads
\begin{eqnarray}
  \label{A-transform}
A_\parallel &=& \frac{1}{1 + \varepsilon R}\, \Big[
  \cos\theta\, \sqrt{1-\varepsilon^2 \phantom{(\!\!}}\, A_1
 + \sin\theta \sqrt{2 \varepsilon (1-\varepsilon)}\; A_2\, \Big] ,
\nonumber \\
A_\perp &=& \frac{1}{1 + \varepsilon R}\, \Big[
  \cos\theta \sqrt{2\varepsilon (1-\varepsilon)}\; A_2
 - \sin\theta \sqrt{1-\varepsilon^2 \phantom{)\!\!}}\, A_1\, \Big] .
\end{eqnarray}
Comparing with (\ref{incl-long}) and (\ref{incl-transv}) we identify
\begin{equation}
A_1 = \frac{\sigma_{++}^{++} - \sigma_{++}^{--}}{
            \sigma_{++}^{++} + \sigma_{++}^{--}} ,
\qquad\qquad
A_2 = \frac{\sqrt{2}\, \sigma_{+0}^{+-}}{
      \sigma_{++}^{++} + \sigma_{++}^{--}} .
\end{equation}
The factors $(1+\varepsilon R)^{-1}= \sigma_T /(\sigma_T +
\varepsilon\sigma_L)$ in (\ref{A-transform}) reflect that the
$\gamma^* p$ asymmetries are defined with respect to the transverse
cross section $\sigma_T$.  Positivity of the matrix (\ref{bigmatrix})
implies $|\, \sigma_{+0}^{+-}| \le (\sigma_{00}^{++}\,
\sigma_{++}^{++})^{1/2}$ and we thus recover the bound
\begin{equation}
| A_2 | \le \sqrt{\half (1+A_1)\, R}
\end{equation}
derived in \cite{Soffer:2000zd}.


\section{Integrals for exclusive meson production}
\label{app:mesons}

For definiteness we give here the convolution integrals appearing in
(\ref{meson-natural}) and (\ref{meson-unnatural}) for $\ell p\to \ell
\rho^0 p$ and for $\ell p\to \ell \pi^+ n$.  Results for other
channels can be found in
\cite{Goeke:2001tz,Diehl:2003ny,Diehl:2003qa}.  To leading order in
$\alpha_s$ one has \\ \\
\begin{eqnarray}
  \label{meson-convolutions}
\mathcal{H}_{\rho^0} &=& \frac{4\pi \alpha_s}{9}\,
  \frac{f_\rho}{ \sqrt{2}} \int_0^1 dz\, \frac{\phi_\rho(z)}{z(1-z)}\,
  \int_{-1}^1 dx\, \Bigg[ 
    \frac{1}{\xi-x-i\varepsilon} - \frac{1}{\xi+x-i\varepsilon} \Bigg]
\\[0.1em]
 && \hspace{13.4em} {}\times
 \Bigg[ \frac{2}{3} H^u(x,\xi,t) + \frac{1}{3} H^d(x,\xi,t) 
      + \frac{3}{8} \frac{H^g(x,\xi,t)}{x} \Bigg] ,
\nonumber \\[0.3em]
\tilde{\mathcal{H}}_{\pi^+} &=& \frac{4\pi \alpha_s}{9} f_\pi
  \int_0^1 dz\, \frac{\phi_\pi(z)}{z(1-z)}\,
  \int_{-1}^1 dx\, \Bigg[ 
    \frac{2}{3}\, \frac{1}{\xi-x-i\varepsilon} 
  + \frac{1}{3}\, \frac{1}{\xi+x-i\varepsilon} \Bigg]
 \Big[ \tilde{H}^u(x,\xi,t) - \tilde{H}^d(x,\xi,t)  \Big]
\nonumber
\end{eqnarray}
with the meson decay constants $f_\rho \approx 209$~MeV and $f_\pi
\approx 131$~MeV and the respective light-cone distribution amplitudes
normalized as $\int_0^1 dz\, \phi(z) = 1$.  Our definitions of GPDs
are such that for $\xi=0$, $t=0$ and $x>0$ they are related to the
usual parton densities in the proton as $H^q(x,0,0) = q(x)$,
$H^g(x,0,0) = x g(x)$ and $\tilde{H}^q(x,0,0) = \Delta q(x)$
\cite{Diehl:2003ny}.  The convolutions $\mathcal{E}_{\rho^0}$ and
$\tilde{\mathcal{E}}_{\pi^+}$ are obtained from
(\ref{meson-convolutions}) by replacing $H$ with $E$ and $\tilde{H}$
with $\tilde{E}$.


\end{document}